\pdfoutput=1
\documentclass[a4paper,11pt]{article}

\usepackage[medium]{titlesec}
\usepackage{subcaption}
\usepackage{graphicx} 
\usepackage{booktabs} 
\usepackage{mathrsfs} 
\usepackage{bm} 
\usepackage{amssymb} 
\usepackage{amsmath}
\usepackage{amsthm}
\newtheorem*{remark}{Remark}
\usepackage{slashed}
\usepackage{hyperref} 
\usepackage{graphicx} 
\usepackage{xcolor} 
\usepackage{comment} 
\usepackage{cite} 
\usepackage{tabularx} 
\usepackage{array} %
\usepackage{multirow} 
\hypersetup{
    colorlinks=true,
    linkcolor=blue,
}

\usepackage[hmargin=.7in,vmargin=1.1in]{geometry}
\usepackage[utf8]{inputenc} 
\usepackage{indentfirst}
\graphicspath{{figs/}}

\newcommand{\black}{\color{black}}

\newcommand{\bge}{\begin{equation}}
\newcommand{\ede}{\end{equation}}
\newcommand{\bga}{\begin{align}}
\newcommand{\eda}{\end{align}}
\def\bgp{\begin{pmatrix}}
\def\edp{\end{pmatrix}}
\def\bgs{\begin{subequations}}
\def\eds{\end{subequations}}

\def\to{\rightarrow}

\def\ii{\mathrm{i}}

\begin{document}

\title{Axion Isocurvature Collider}

\author{
Shiyun Lu$^a$
\\[3mm]
\normalsize{$^a$~\emph{Department of Physics, The Hong Kong University of Science and Technology,}}\\
\normalsize{\emph{Clear Water Bay, Kowloon, Hong Kong, P.R.China}}\\
\normalsize{\emph{Jockey Club Institute for Advanced Study, The Hong Kong University of Science and Technology,}}\\
\normalsize{\emph{Clear Water Bay, Kowloon, Hong Kong, P.R.China}}\\
}

\maketitle

\vspace{2cm}

\begin{abstract}

Cosmological colliders can preserve information from interactions at very high energy scale, and imprint them on cosmological observables. Taking the squeezed limit of cosmological perturbation bispectrum, information of the intermediate particle can be directly extracted from observations such as cosmological microwave background (CMB). Thus cosmological colliders can be powerful and promising tools to test theoretical models. In this paper, we study extremely light axions (including QCD axions and axion-like-particles), and consider them constituting cold dark matter (CDM) at late times. We are interested in inflationary isocurvature modes by such axions, and try to figure out how axion perturbations can behave as isocurvature colliders.  We work out an example where the intermediate particle is a boson, and show that, in the squeezed limit, it is possible to provide a clock signal of significant amplitudes, with a characteristic angular dependence. This provides a channel to contribute and analyze clock signals of isocurvature bispectrum, which we may hopefully see in future experiments.

\end{abstract}

\newpage

\black
\tableofcontents

\newpage
\section{Introduction}
In the standard model of cosmology, the universe went though the early epoch with extremely high energy, and there was an exponentially expanding period at the very beginning, driven by inflaton. Quantum fluctuations during inflation perturbed the universe, usually described by gauge invariant curvature perturbations $\zeta$, and enter observations such as cosmological microwave background (CMB) and large scale structure (LSS). For inflation driven by a single field, the primordial perturbation is adiabatic, and the curvature perturbation $\zeta_\phi\simeq-\frac{H}{\dot\phi}\delta\phi$. While in multi-field inflation, the existence of other fields evolving differently at late time can cause isocurvature perturbations, even when the total density unchanged. The entropy $S$ is used to describe the differences between species $S_{ij}=3(\zeta_i-\zeta_j)$, where the curvature perturbation of species $i$ is defined by $\zeta_i=-\Psi-H\delta\rho_i/\dot{\bar\rho}_i$ \footnote{The expression here omits the fraction of species $i$.}. 
\\ \indent
Cosmological perturbations interacting during inflation can serve as colliders at an energy scale much higher than possibly reached at labs. These perturbations leave information on the CMB. Such cosmological colliders have been studied in\cite{Chen:2009we, Chen:2009zp, Baumann:2011nk, Noumi:2012vr, Arkani-Hamed:2015bza, Lee:2016vti, Baumann:2017jvh,
Assassi:2012zq, Sefusatti:2012ye, Norena:2012yi, Emami:2013lma, Liu:2015tza, Delacretaz:2016nhw, Kehagias:2015jha, Chen:2018xck, Dimastrogiovanni:2015pla, Schmidt:2015xka, Chen:2015lza, Delacretaz:2015edn, Bonga:2015urq, Flauger:2016idt, Chen:2016nrs, Chen:2016uwp, Meerburg:2016zdz, Chen:2016hrz, An:2017hlx, Tong:2017iat, Iyer:2017qzw, An:2017rwo, Kumar:2017ecc, Riquelme:2017bxt, Saito:2018omt, Cabass:2018roz, Wang:2018tbf, Dimastrogiovanni:2018uqy, Bordin:2018pca, Chua:2018dqh, Arkani-Hamed:2018kmz, Kumar:2018jxz, Goon:2018fyu, Wu:2018lmx, Li:2019ves, McAneny:2019epy, Kim:2019wjo, Sleight:2019mgd, Biagetti:2019bnp, Sleight:2019hfp, Welling:2019bib, Alexander:2019vtb, Lu:2019tjj, Hook:2019zxa, Hook:2019vcn, Kumar:2019ebj, Liu:2019fag, ScheihingHitschfeld:2019tzr, Wang:2019gbi, Baumann:2019oyu, Wang:2019gok, Wang:2020uic, Li:2020xwr, Wang:2020ioa, Baumann:2020dch, Kogai:2020vzz, Bodas:2020yho, Aoki:2020zbj, Maru:2021ezc, Kim:2021pbr
}. That is, during inflation, the coupling of intermediate particle $\sigma$ and the inflaton $\phi$ can provide a channel to correlate inflatons, leaving imprints on curvature correlation functions. By some analysis, the spectra in momentum space can give us information of the intermediate particle $\sigma$, such as its mass and spin. A very straightforward correspondence has been shown. The three point function of inflaton fluctuations in momentum space has a general form like \cite{Arkani-Hamed:2015bza}
\begin{align}
	\langle\delta\phi(\bm{k_1})\delta\phi(\bm{k_2})\delta\phi(\bm{k_3})\rangle\sim(2\pi)^3\delta^{(3)}(\bm{k_1}+\bm{k_2}+\bm{k_3})\sum P_s(\cos\theta)A(\mu_m)\left(\frac{k_3}{k_1}\right)^{\Delta(\mu_m)}\ \ \text{for }k_3\ll k_1\simeq k_2 ~,
\label{eqt.3squeeze}
\end{align}
in the squeezed limit, where $P_s$ is the Legendre polynomials, $s$ and $\mu_m$ are dependent on the spin and the mass of the interacting particle, respectively, 
\begin{figure*}[h]
	\centering
	\includegraphics[width=0.3\textwidth]{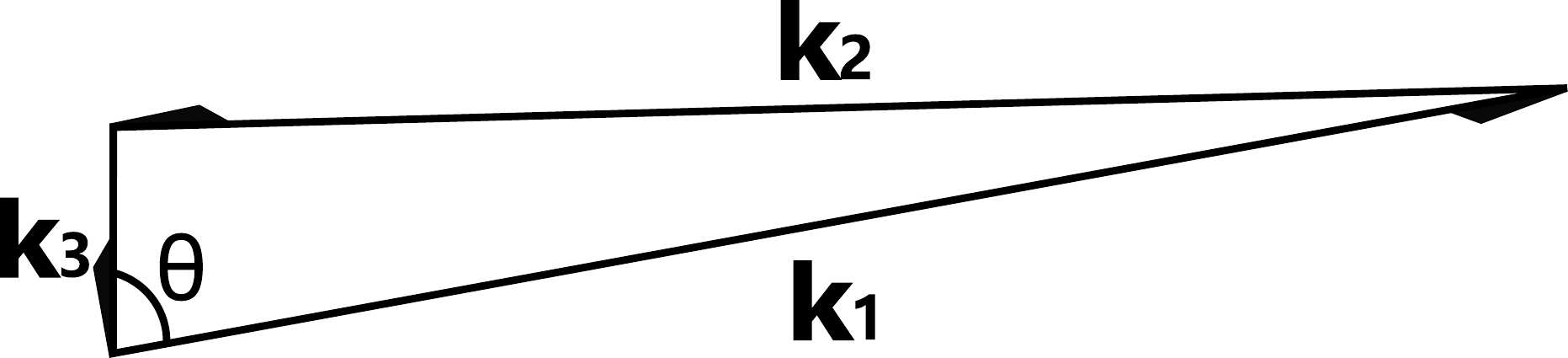}
\end{figure*}
\\ and $\theta$ is the angle in the momentum diagram shown above. The inflaton fluctuations are converted to cosmological adiabatic perturbations, thus the characteristic power-law (from the real part of $\Delta$) or periodic (from the imaginary part of $\Delta$) clock signals in \eqref{eqt.3squeeze} will be reflected on cosmological microwave background (CMB). Such inflaton colliders have been considered for detecting standard model (SM) physics \cite{Chen:2016nrs,Chen:2016uwp,Chen:2016hrz,Kumar:2017ecc}, neutrino physics \cite{Chen:2018xck} and some beyond-SM new physics \cite{Kumar:2018jxz}.
\\ \indent
While the adiabatic colliders have been well-studied, we turn our interests to isocurvature colliders. Isocurvature perturbations can hopefully give significant contribution to bispectrum. As these perturbations can be affected by post-inflationary evolution, they are rather model-dependent \cite{Baumann:2009ds}. We have already discussed Higgs as isocurvature collider in our previous work \cite{Lu:2019tjj}. A curvaton case has also been studied in \cite{Kumar:2019ebj}.
\\ \indent
A field that was subdominant during inflation, long-lived to become more dominant and leave relic density, can be a natural consideration for isocurvature collider. In this paper, we study exremely light axions, which has large parameter space and different model constructions, and can be good candidate to constitute dark matter (DM) \cite{Preskill:1982cy,Abbott:1982af,Dine:1982ah}. Axions are weakly coupled to standard model (SM) particles, thus can be introduced freely without large corrections to previous theories. As a pseudoscalar field, it can also record the spin information of the interacting particle, through the chiral asymmetric interaction, such as $c_\gamma\chi F\tilde{F}/f_a$ for bosons or ${c_\psi}\bar\psi\slashed{\partial}\chi\gamma_5\psi/({f_a m_\psi})$ for fermions, and gives angular $\theta$-dependent signals in the bispectrum.
\\ \indent
Axion models were first proposed to solve the strong CP problem. That is, the neutron electric dipole momentum (eDM) measurement \cite{Graner:2016ses} gives the constrain $\theta_{\text{QCD}}\lesssim 10^{-10}$, causing a fine-tuning problem. By non-perturbative effects like instantons \cite{Vafa:1984xg}, QCD axions provide a potential which reach the minimum when the anomaly angle is zero \cite{Peccei:1977hh,Weinberg:1977ma,Wilczek:1977pj}, thus can solve the problem. Similar fields can also arise in effective field theory of string theories at low energy scale and predicts axion-like-particles (ALPs) \cite{Svrcek:2006yi}. The topological structure of a string model can give rise to hundreds of ALPs with masses separated in a hierarchy manner, and this is called an ``axiverse'' \cite{Arvanitaki:2009fg, Acharya:2010zx, Cicoli:2012sz}. We will discuss both QCD axions and ALPs in our paper.
\\ \indent
We show that, the axion collider can leave significant angular-dependent clock signals on cosmological isocurvature bispectrum, within the allowed parameter space. We study in detail about the case with bosonic intermediate particle $Z$, and give clock signals in the squeezed limit, with the possibility of obvious periodic properties and large contribution to the non-Gaussianity of isocurvature mode. We can see that, ALP models have a variety of mechanisms, and the constrains are much loosened. The amplitudes of our signals are generally suppressed when $\mu=\sqrt{(m_Z/H_I)^2-1/4}$ gets larger, $(H_I/f_a)$ gets smaller, or $c=c_\gamma \dot\chi_0/(f_a H_I)$ gets smaller. Note that, although the coupling constant $c_\gamma$ can usually be $\mathcal{O}(1)$, light DM axions with $m_a\ll H_I$ roll too slow during inflation, making $c=c_\gamma \dot\chi_0/(f_a H_I)$ extremely small, and all bispectrum signals vanish. Mechanisms to allow dynamic ALP mass may solve this problem, such as discussed in \cite{Marsh:2011gr}. With these mechanisms, our expected large signals can be produced by axion models of reasonable parameters without fine-tuning problems.
\\ \indent
This paper is organized as following: In Sec.~\ref{sec.axion}, we introduce the axion model, and discuss the evolution during cosmological epochs; In Sec.~\ref{sec.fluc}, we study the fluctuations and give the mode functions during inflation; In Sec.~\ref{sec.nonGau}, we work out the 1-loop bispectrum contributed by axion-gauge boson interaction, and approximate the results under reasonable assumptions; In Sec.~\ref{sec.bi}, we estimate the bispectrum signal under constrains.
\section{Axion models and cosmological implications}\label{sec.axion}
\subsection{Axion Models}
\paragraph{QCD axions}
Quantum chromodynamics (QCD) naturally contains a term \cite{Marsh:2015xka}
\begin{align}
	\frac{\theta_{\text{QCD}}}{32\pi^2}\text{Tr}(F_{\mu\nu}\tilde{F}^{\mu\nu}) ~,
\end{align}
in its Lagrangian. This term is CP-violating and the corresponding neutron electron dipole momentum (eDM) is $d_n\approx 3.6\times 10^{-16} \theta_{\text{QCD}}\ e\ \text{cm}$, which is constrained by experiments, and thus gives the bound $\theta_{\text{QCD}}\lesssim 10^{-10}$. To solve this fine-tuning problem, a pseudoscalar field known as the QCD axion is introduced. There are several models \footnote{PQWW (additional complex scalar field as a second Higgs doublet), KSVZ (additional heavy quark doublet, while the PQ scalar field as a SM singlet) and DFSZ (additional Higgs doublet, while the PQ scalar field as a SM singlet) models. See \cite{Hook:2018dlk} for details.} to achieve this purpose through a complex PQ scalar $\Phi$, which gets spontaneously broken at the scale of $f_a$.
At classical level, the Lagrangians of these models are invariant under a PQ rotation
\begin{align}
	\psi \to e^{\ii \mathcal{Q}_{\text{PQ},i} \chi/f_a}\psi ~,
\end{align}
for field $\psi$, where $\mathcal{Q}$ is the PQ charge, and need to add a $\gamma_5$ if $\psi$ is a spinor. However, taking the anomaly into consideration, a new term appears,
\begin{align}
	\mathcal{L} \to \mathcal{L} + \frac{1}{32\pi^2}\frac{\mathcal{C}\chi}{f_a}\text{Tr}(F_{\mu\nu}\tilde{F}^{\mu\nu}) ~,
\end{align}
where
\begin{align}
	\mathcal{C}\delta_{ab}=2\text{Tr}\left(\mathcal{Q}_{PQ,i}T_aT_b\right)
\end{align}
is the color anomaly of PQ symmetry, $T_a$ is the generator of $SU(3)$. Due to the requirement of periodic properties by $\theta$-vacua \cite{coleman_1985}, $\mathcal{C}$ should be an integer $\mathcal{C}=N_{DW}$. Thus the transformation can be effectively seen as $\theta_{\text{QCD}} \to \theta_{\text{QCD}}+ N_{DW} \chi/f_a$. If we treat $\chi$ as a field, the overall angular ($\theta_{\text{QCD}}+ N_{DW} \chi/f_a$) becomes dynamic and can naturally go to zero, given an axion potential like \footnote{Here the constant $\theta_{\text{QCD}}$ is absorbed into the axion field.}
\begin{align}
	V(\chi)=\Lambda_a^4(1-\cos({N}_{\text{DW}}\chi/f_a)) ~,
\label{eqt.apotential}
\end{align}
that reaches its minimum at zero field value. Such a potential is from the vacuum energy, produced by non-perturbative effects like instantons. The pseudo Nambu-Goldstone boson $\chi$ here is the QCD axion.
In this way, the strong CP problem can be solved.
\\
\indent
The domain wall number $\mathcal{N}_{\text{DW}}$ is a positive integer usually taken to be 1, and $\Lambda_a^4\simeq m_u\Lambda_{\text{QCD}}^3$\footnote{Here the expression is at the low temperature limit $T\to 0$, actually this is temperature-dependent, which is significant at high temperature.} for QCD axions. 
The mass of axions produced by QCD instantons,
\begin{align}
	m_{a,\text{QCD}}\approx 6\times 10^{-6}\ \text{eV}\left(\frac{10^{12}\ \text{GeV}}{f_a/\mathcal{C}}\right) ~,
\label{eqt.malowT}
\end{align}
is model independent \cite{Wilczek:1977pj, Weinberg:1977ma}.
\paragraph{ALPs}
The above axion model can be extended to more general pseudo Nambu-Goldstone bosons (pNGBs) including axion-like particles (ALPs). This is often achieved by some string models, the axion produced by which also require non-perturbative effects. The instanton action contribute an exponential factor to the axion potential \cite{Callan:1977gz},
\begin{align}
	e^{-S_{\text{inst.}}}=e^{-8\pi^2/g_i^2} ~,
\end{align}
where $g_i$ is the coupling of the gauge group. In high-dimensional theories, each way of extra dimensions wrapped corresponds to a $g_i$, and $g_i^2\propto 1/\sigma_i$, where $\sigma_i$ is the modulus. Thus finally \cite{Marsh:2015xka},
\begin{align}
	\Lambda_a^4=M_{pl}^2 m_{\text{SUSY}}^2 e^{-\#\sigma_i} ~,
\end{align}
with a dependence on modulus, and the axion mass
\begin{align}
	m_a^2\simeq \Lambda_a^4/f_a^2 ~.
\end{align}
As the modulus can take a large range of values in different models, this effectively makes the axion mass $m_a$ a ``free parameter''. The plenitude of axions is called an ``axiverse'' \cite{Arvanitaki:2009fg}.
\paragraph{Scales}
When the energy scale goes down to $f_a$, the PQ symmetry of scalar field $\Phi$ is spontaneously broken. Then when the temperature $T\simeq \Lambda_{\text{QCD}}$, non-perturbative effects such as instantons are switched on. For QCD axions, $\Lambda_{\text{QCD}}\approx 200\ \text{MeV}$, and decay constant $f_a: 10^9\sim 10^{17}~\text{GeV}$, where the upper bound and lower bound are from black hole superradiance (BHSR) and supernova cooling, respectively \cite{Marsh:2015xka}. Thus the QCD axion mass range is $m_{a,\text{QCD}}=4\times 10^{-10}\sim 4\times 10^{-1}~\text{eV}$. In string models, $f_a$ are typically of GUT scale $\lesssim 10^{16}~\text{GeV}$ and can possibly reach lower value as $10^{10\sim 12}~\text{GeV}$. The axions we are considering can have a large mass range from $10^{-33}~\text{eV}$ to $10^{-3}~\text{eV}$, where the lower bound is from the Hubble parameter today $H_0\approx 10^{-33}~\text{eV}$ \footnote{Cause the axion with mass lower than this do not start oscillate till today.}. The axions with $10^{-33}~\text{eV}\lesssim m_a\lesssim 10^{-18}~\text{eV}$ are called ultralight axions (ULAs).
\\ \indent
The Hubble parameter during inflation is $H_I= 8\times 10^{13}\sqrt{r_T/0.1}~\text{GeV}$ \cite{Enqvist:2017kzh}, where the tensor-to-scalar ratio $r_T<0.07$ is constrained by Planck and BICEP2 \cite{Ade:2018gkx}.
\paragraph{Interaction with standard model}
The couplings of QCD axions are computed in \cite{Srednicki:1985xd}. 
No matter what specific model is, axions (including ALPs) interact with SM particles as following,
\begin{align}
	\frac{g_{\chi\psi}}{2m_\psi}\partial_\mu \chi(\bar{\psi}\gamma^\mu\gamma_5\psi) ~,
\end{align}
for fermion $\psi$ coupling, and
\begin{align}
	-\frac{1}{4}g_{\chi Z}\chi F_{\mu\nu}\tilde{F}^{\mu\nu} ~,
\label{eqt.interactZ}
\end{align}
for boson $Z$ coupling. In this paper, we will focus on the bosonic interaction. The coupling constants for QCD axions and some specific ALP models can be computed \cite{Dias:2014osa,Kim:2015yna}. For example, QCD axions couple to photons,
\begin{align}
	g_{\chi\gamma}= \frac{\alpha_{\text{EM}}}{2\pi f_a}\mathcal{N}_{DW}c_{\chi\gamma} ~,
\label{eqt.cog}
\end{align}
where $\alpha_{\text{EM}}=1/137$ is the EM coupling constant, and $c_{\chi\gamma}$ is model dependent, usually of $\mathcal{O}(1)$. The large decay constant $f_a$ makes a weak coupling to SM, and also a small mass of the axion, thus makes axion a good candidate for DM. 
\\ \indent
For generic ALP models, the coupling constants are taken to be free parameters, and can be much less than those of QCD axions. Therefore, in this paper, we do not care about the explicit expression of the coupling constants and treat them as free parameters.
\subsection{Cosmological axions}
\begin{figure*}[t]
	\centering
	\includegraphics[width=0.45\textwidth]{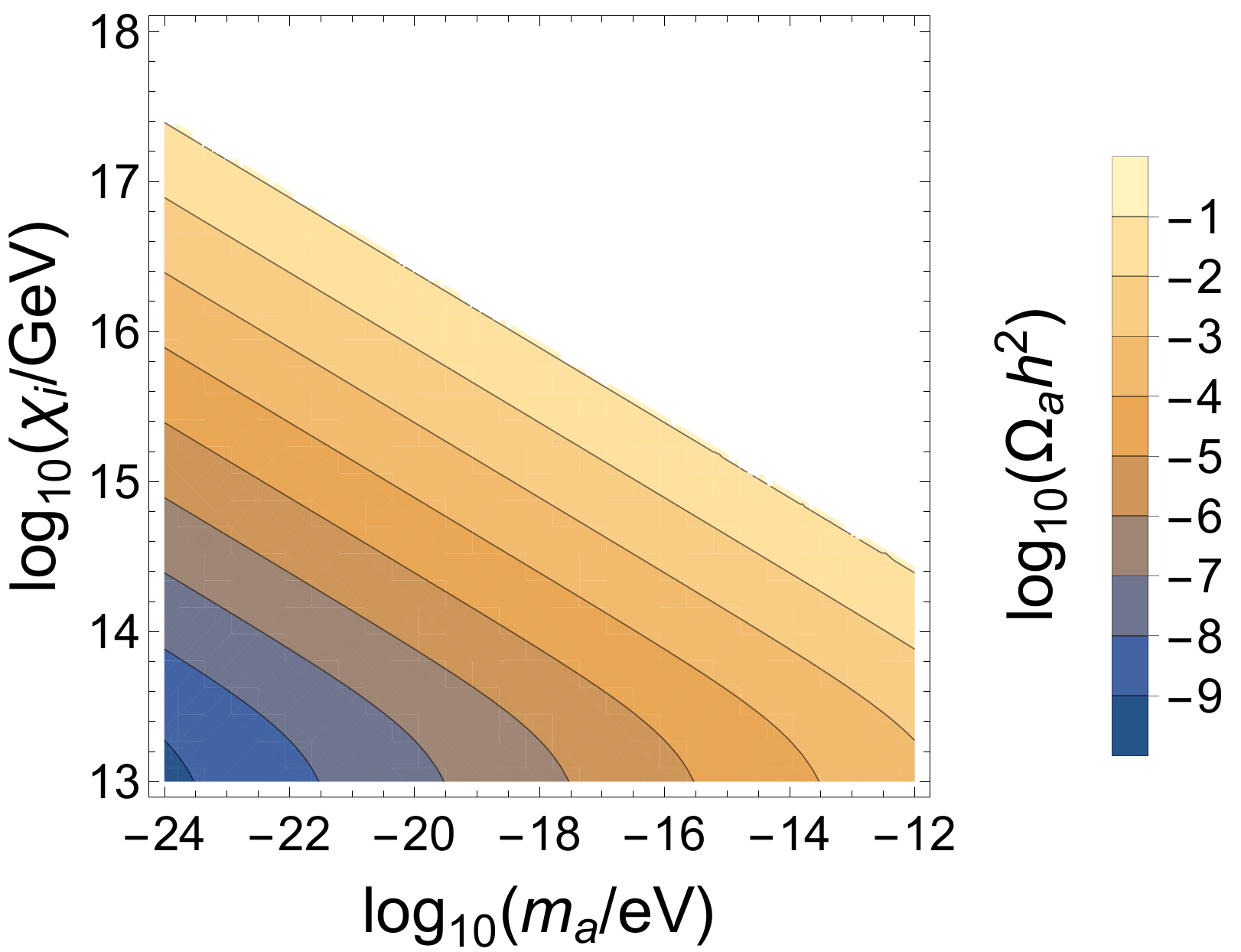}
	\caption{Relic density of ALP dark matter produced through misalignment, the oscillation of which began during the RD epoch, with $H_I=10^{14}~\text{GeV}$, $H_0=100h$ km$*\text{s}^{-1}\text{Mpc}^{-1}$ ($h$=0.67), $M_{\text{pl}}=2.44\times 10^{18}~\text{GeV}$, $\Omega_rh^2=4.18\times 10^{-5}$. The region shows axion relic density when it does not excess the overall cold dark matter (CDM) density $\Omega_d h^2=0.12$. The initial misalignment angle $\theta_i=\chi_i/f_a$. Reproduced from \cite{Marsh:2015xka}. It can be seen that, to constitute a non-negligible fraction of $\Omega_d$, we should have decay constant $f_a\gtrsim \chi_i>10^{14}~\text{GeV}$, generally larger than inflationary Hubble constant, satisfying PQ broken scenario condition. }
	\label{fig.alpdm}
\end{figure*}

\begin{figure*}[t]
	\centering
	\begin{subfigure}{0.45\textwidth}
	\caption{$H_I=2\pi\times 10^9 \text{GeV}$\hspace*{20pt}}
	\includegraphics[width=\textwidth]{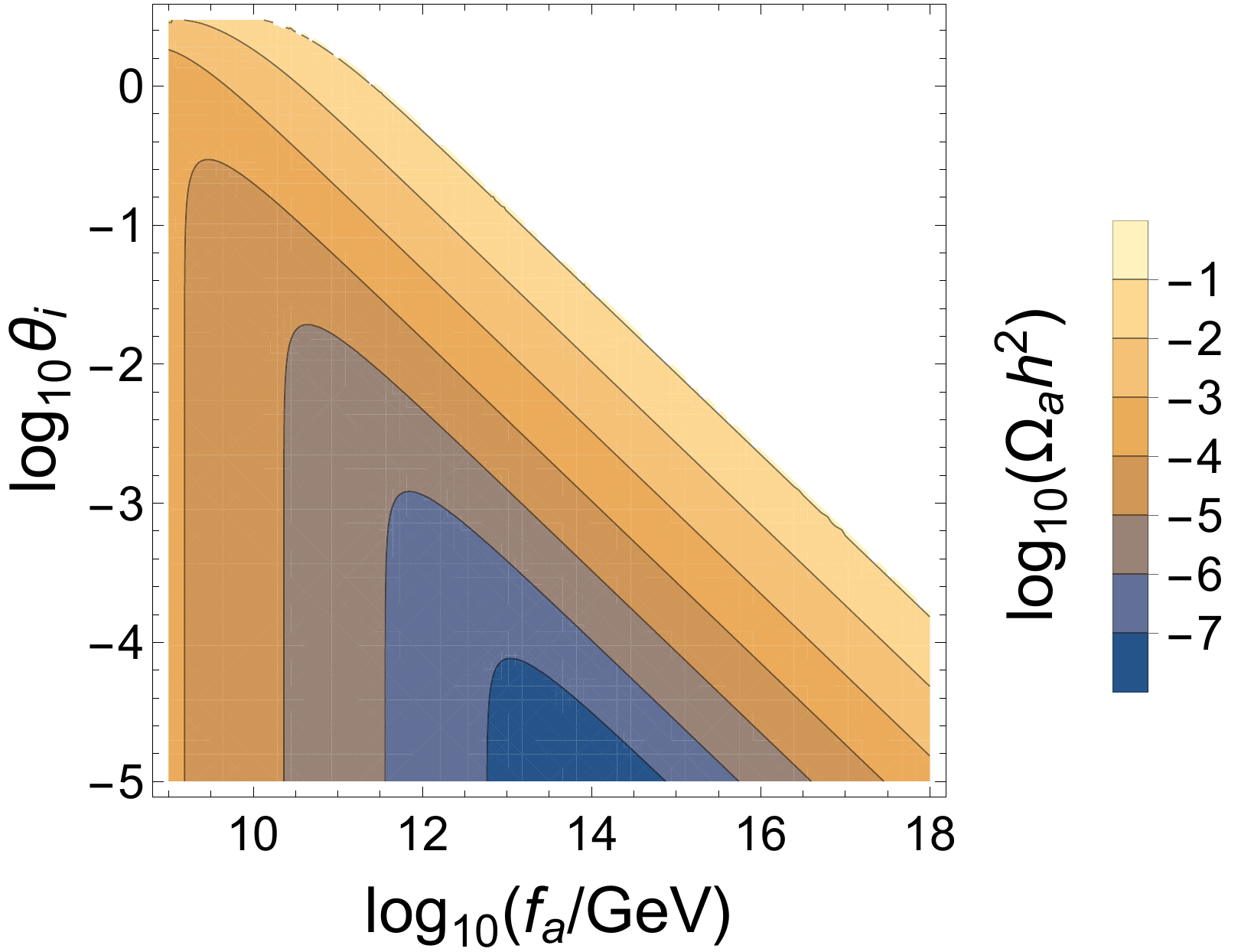}
	\end{subfigure}
\hspace*{30pt}
	\begin{subfigure}{0.45\textwidth}
	\caption{$H_I=10^{14} \text{GeV}$\hspace*{30pt}}
	\includegraphics[width=\textwidth]{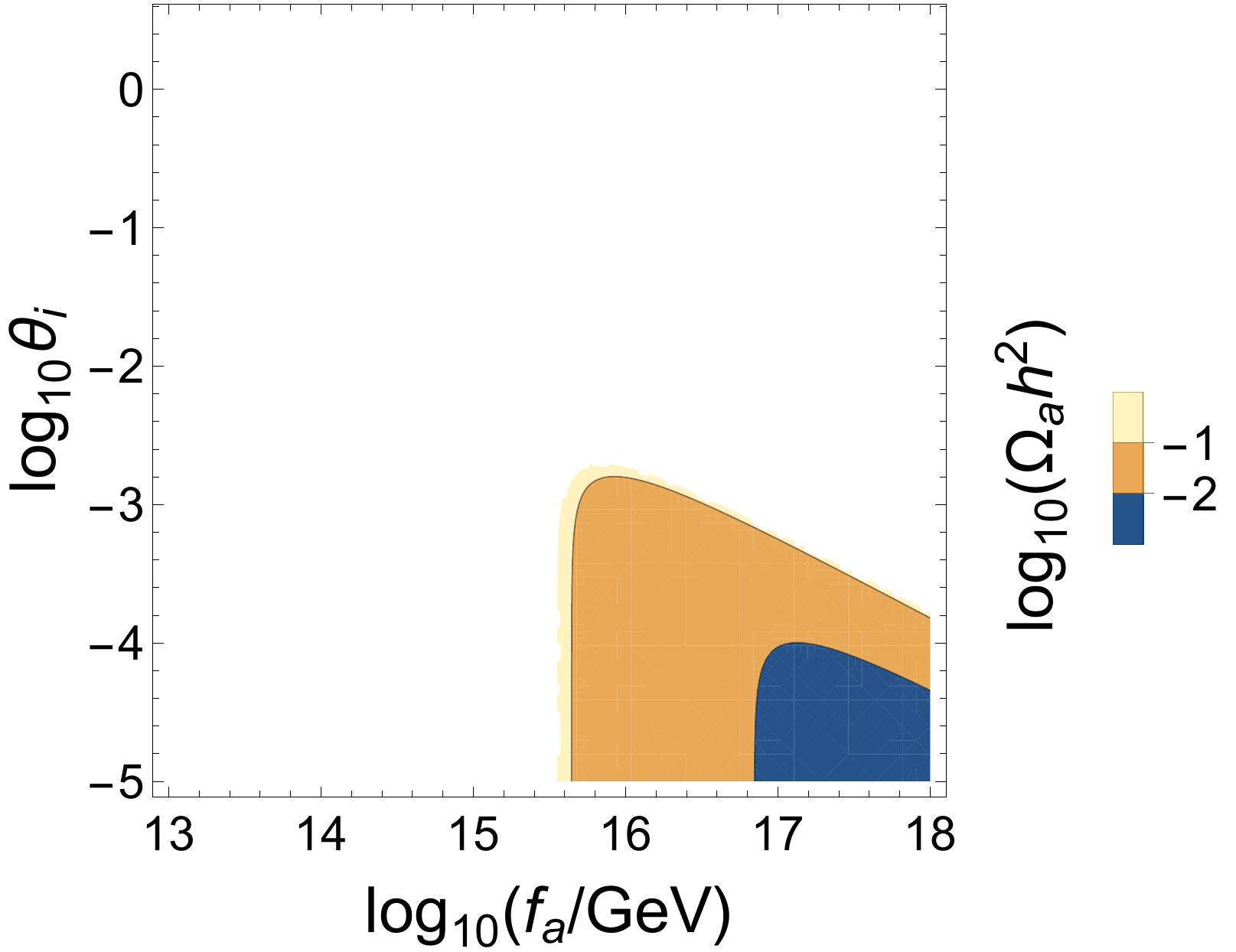}
	\end{subfigure}
	\caption{Relic density of QCD axion dark matter produced through misalignment in PQ broken scenario, and $f_a<10^{18}~\text{GeV}$. The region shows axion relic density when it does not excess the overall cold dark matter (CDM) density $\Omega_d h^2=0.12$. The left figure is for a inflationary scale $H_I=2\pi\times 10^9~$GeV. While the right one is for $H_I=10^{14}~$GeV, when only fine-tuning $\theta_i$ is allowed due to abundance constrain. Reproduced from \cite{Marsh:2015xka}.}
	\label{fig.qcdadm}
\end{figure*}

\begin{figure*}
	\centering
	\includegraphics[width=0.45\textwidth]{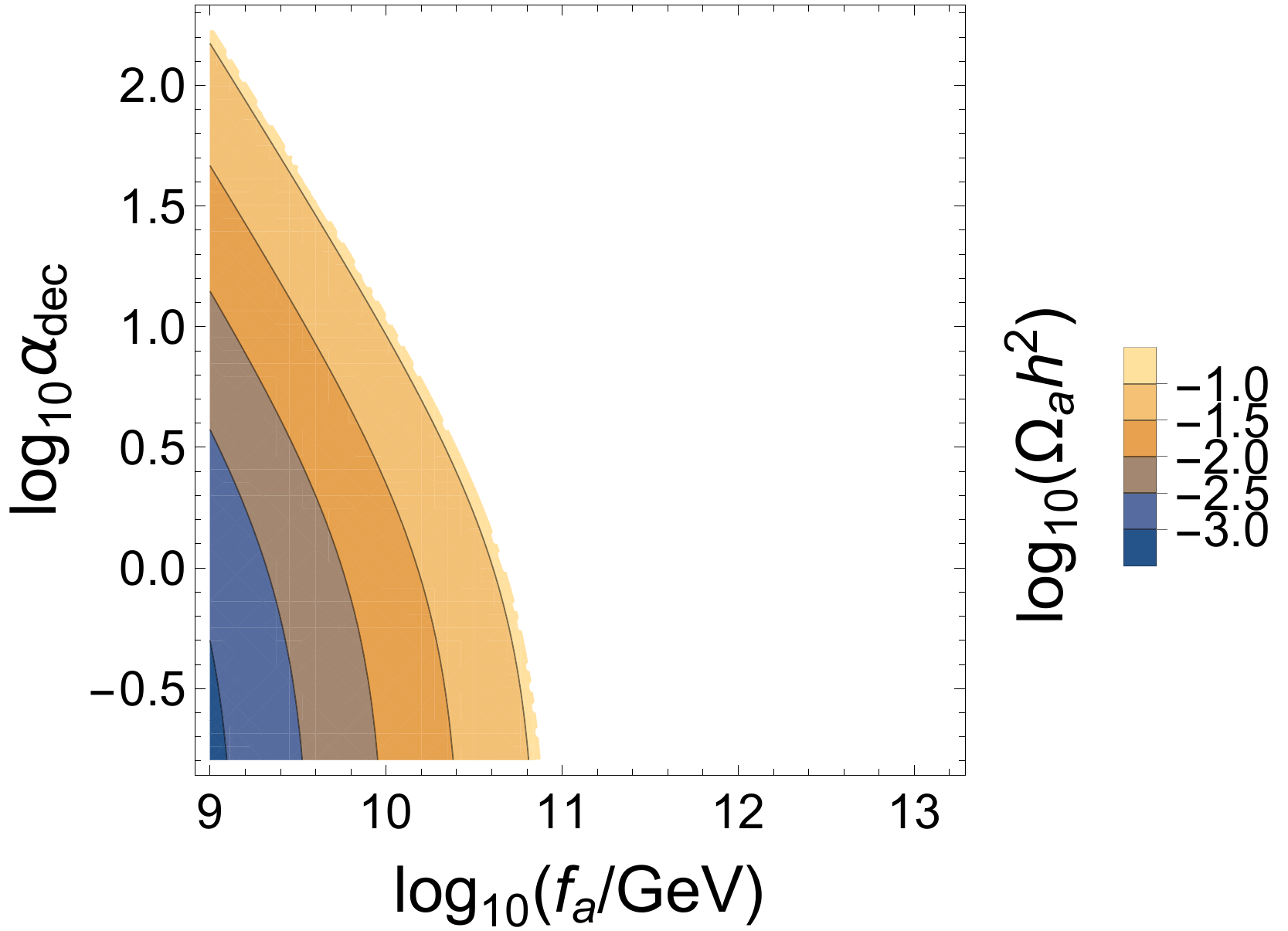}
	\caption{Relic density of QCD axion dark matter produced through misalignment and topological defects decay, in PQ unbroken scenario. The region shows axion relic density when it does not excess the overall cold dark matter (CDM) density $\Omega_d h^2=0.12$. Reproduced from \cite{Marsh:2015xka}.}
	\label{fig.qcdadmu}
\end{figure*}

\paragraph{During inflation}
The axion field gets an initial value when PQ symmetry gets broken at the energy scale $f_a$. As there is no special value different from others, this can be treated as a random value, $\theta_{i}=\chi_{i}/f_a$ taken from $[-\pi,\pi]$ randomly for each patch of the universe, as our axion potential $V(\chi)$ in \eqref{eqt.apotential} is periodic. Generally this random initial value by symmetry breaking do not happen to locate at the minimum of vacuum potential $V(\theta)$, this is called ``misalignment'', and the axion background value $\theta_0$ rolling back to the minimum is called ``realignment''. 
\\ \indent
It is important to distinguish the cases when PQ symmetry breaking happens during of after inflation. Cause the rapid expansion during inflation can dilute away the patches with different $\theta_{i}$ in the PQ broken scenario.
\begin{itemize}
	\item If $f_a\lesssim H_I/2\pi$, PQ symmetry is broken after inflation, so in the late time universe, taking average of all patches we have
\end{itemize}
\bga
	\langle\chi_i^2\rangle=f_a^2\pi^2/3 ~,
\end{align}
\begin{itemize}
	\item If $f_a\gtrsim H_I/2\pi$, PQ symmetry is broken before or during inflation. So in our universe patch within horizon, we have
\end{itemize}
\bga
	\langle\chi_i^2\rangle=f_a^2\theta_{i}^2+\langle{\delta\chi}^2\rangle ~,
\end{align}
where $\theta_{i}$ is the initial value in our patch, which is undetermined and be treated as a free parameter. $\delta\chi=H_I/(2\pi)$ is the isocurvature perturbation of axion field as the quantum effect within horizon. In our case that axions compose DM, we have $f_a\gtrsim H_I/2\pi$ most of the time, otherwise there will be over production. More details will be shown below.
\paragraph{Background evolution (after inflation)}
For $\chi/f_a=\theta_{a}\ll 1$, the potential from non-perturbative effect is approximately $V(\chi)\simeq \frac{1}{2}m_a^2\chi^2$, so we can write the equation of motion of axion background \footnote{If we take the back reaction from particle production into consideration, an effective dissipation $\Gamma$ should be added to $3H$.}
\begin{align}
	\ddot\chi_0+(3H+\Gamma)\dot\chi_0+m_a^2\chi_0=0 ~,
\label{eqt.chieom}
\end{align}
and the density and pressure of axion 
\begin{align}
	\bar p_a=\frac{1}{2}\dot\chi_0^2-\frac{1}{2}m_a^2\chi_0^2 ~,\\
	\bar\rho_a=\frac{1}{2}\dot\chi_0^2+\frac{1}{2}m_a^2\chi_0^2 ~.
\end{align}
\indent
When $H\gg m_a$, the large friction term will prevent $\chi_0$ background value from any significant change, with a very slow $\dot\chi_0\simeq -\frac{m_a^2}{3H}\chi_0$. This is ``over damped'', at which we can treat $\chi_0$ almost frozen as $\chi_0\simeq f_a\theta_{i}$. As $\omega\simeq -1$ under this condition, the axion behaves like dark energy, and its energy density remains constant.\\
\indent
When energy scale goes down to $H\approx m_a$, the axion background begins oscillation and quickly goes to the asymptotic state. We can solve \eqref{eqt.chieom} in a radiation-dominant (RD) or matter-dominant (MD) universe, which has the scale factor $a\propto t^p$
\begin{align}
	\chi_0(t)=a^{-3/2}(t/t_i)^{1/2}[C_1J_{(3p-1)/2}(m_a t)+C_2Y_{(3p-1)/2}(m_a t)] ~,
\end{align}
where $J$ and $Y$ are Bessel functions, and $t_i$ is the time when axion exited ``frozen'' status and began oscillation. 
However, in a universe where matter and radiation both have significant contributions, there nay bot be an analytical solution. But as long as $H\ll m_a$, we can use the WKB approximation,
\begin{align}
	\chi_0(t)=\mathcal{A}(t)\cos(m_a t+\mathcal{C}) ~,
\end{align}
where the amplitude $\mathcal{A}(t)$ varies slowly $\dot{\mathcal{A}}(t)/m_a\sim H/m_a\ll 1$. Substituting $\chi_0(t)$ with the approximate solution into \eqref{eqt.chieom}, we can have $\mathcal{A}(t)\sim a^{-3/2}$. Thus, as $H\ll m_a$ a period of time after oscillation begins, the energy density of axion $\rho_a\sim a^{-3}$, axions can be effectively treated as cold dark matter (CDM).\\
\indent
The axion can be treated as $\rho_a=m_a^2 \chi_i^2 /2=\text{const}$ until oscillation begins and $\rho_a(a)=\rho_a(a_{\text{osc}})(a/a_{\text{osc}})^{-3}$ after that, neglecting the transition time. To better fit the result under the short transition time approximation, we need to make a numerical substitution to locate the $a_{\text{osc}}$
\begin{align}
	AH(a_{\text{osc}})=m_a ~,
\end{align}
and this fits well when the coefficient $A$ is taken to be 3 \cite{Hlozek:2014lca}. We can see that, for axion constituting DM, the mass $m_a>H_0\simeq 10^{-33}$ eV Hubble today, or axions lighter than this can constitute dark energy (DE).
\paragraph{Production of axions}
Axions can be produced from several mechanisms \cite{Marsh:2015xka}.
\begin{itemize}
	\item{Decay product of its parent particle. The parent particle $X$ decays to relativistic axions $\chi$. If the decay happens after axions decoupled from SM particles, axions behave as dark radiation (DR). This generally occurs in models with supersymmetry and extra dimensions. Axions produced by moduli $\sigma$ decay survive from before BBN till today. 
}
	\item{Thermal productions from SM radiation. Axions in thermal contact with SM particles lead to thermal relic population of axions. ALPs couple to SM particles more weakly than QCD axions, thus get less thermal productions. For QCD axions, the relic population is negligible if the decay constant $f_a\gtrsim 10^9~\text{GeV}$, as constrained by stellar cooling.}
	\item{Misalignment $m_a^2\chi_i^2/2$. After PQ symmetry is broken, the initial value of the axion field away from the minimum of potential is called ``misalignment''. This also contribute to the relic density of axions, as discussed above. After $H\sim m_a$, the axion starts oscillation and behaves as CDM.}
	\item{Decay product of topological defect. If PQ is unbroken during inflation and thus topological defects are not diluted away. Axions produced by the decay of topological defect are dominated by non-relativistic ones, and contribute to CDM. The simulation gives $\rho_{a,\text{dec}}=\rho_{a,{\text{mis}}}\alpha_{\text{dec}}$, where $\alpha_{\text{dec}}$ has an estimated value ranging from $0.16$ to $186$ \cite{Marsh:2015xka, Hiramatsu:2012gg}.}
\end{itemize}
\indent
We discuss the whole picture for axion production, for ALP and QCD axions, respectively.
For ALPs, the temperature that non-perturbative effect is switched on is much larger than temperature at the late time when oscillation begins, so we can directly take $T\to 0$ limit, that the axion mass has reached the asymptotic value and can be treated as constant. Then we can write the component of axion DM relic density, which oscillated during radiation-dominant (RD) epoch, matter-dominant (MD) epoch or has not oscillated till today
\begin{align}
	\Omega_a\approx \left\{
	\begin{matrix}
		\frac{1}{6}(A^2\Omega_r)^{3/4}(m_a/H_0)^{1/2}\langle \chi_i^2/M_{\text{pl}}^2 \rangle & a_{\text{osc}}<a_{\text{eq}}\\
		\frac{1}{6}A^2\Omega_m \langle \chi_i^2/M_{\text{pl}}^2 \rangle & a_{\text{eq}}<a_{\text{osc}}\lesssim 1\\
		\frac{1}{6}(m_a/H_0)^{2}\langle \chi_i^2/M_{\text{pl}}^2 \rangle & a_{\text{osc}}\gtrsim 1
	\end{matrix}
	\right. ~.
\label{eqt.relica}
\end{align}
where $H_0$ is the Hubble constant today, $a_{\text{eq}}$ is the scale factor at matter-radiation equality (when the redshift $z_{\text{eq}}\approx 3300$ and meanwhile the Hubble constant $H_{\text{eq}}\sim 10^{-27}$~eV), a numerical approximation parameter is taken as $A=3$, and the expectation value of $\langle \chi_i^2\rangle=f_a^2\theta_{i}^2+(H_I/2\pi)^2$. In Fig.~\ref{fig.alpdm}, we show the relic density of ALPs, when PQ was broken during inflation, started oscillation during RD epoch and constitute cold DM today.
\\ \indent
\\ \indent
For QCD axions, we need to consider the change of axion mass, as the non-perturbative effects were switched on at $T\approx \Lambda_{\text{QCD}}\approx 200~\text{MeV}$. More careful consideration for high temperature $T>1~\text{GeV}$ gives \cite{Fox:2004kb}
\begin{align}
	m_a^2(T)=\alpha_a\frac{\Lambda_{\text{QCD}}^3 m_u}{f_a^2}\left(\frac{T}{\Lambda_{\text{QCD}}}\right)^{-n} ~,
\end{align}
while at low temperature $T\lesssim\Lambda_{\text{QCD}}$, the mass is as shown in \eqref{eqt.malowT}. The QCD axions with $f_a<2\times 10^{15}~\text{GeV}$ begins oscillation when $T>1~\text{GeV}$. 
The relic density for $f_a<2\times 10^{15}~\text{GeV}$ is given by \cite{Fox:2004kb}
\begin{align}
	\Omega_a h^2\approx
	2\times 10^4\left(\frac{f_a}{10^{16}~\text{GeV}}\right)^{7/6}\langle\chi_i^2/f_a^2\rangle ~,
\label{eqt.densityQCD}
\end{align}
but this expression can also give good approximation until $f_a<6\times 10^{17}~\text{GeV}$. We use $\eqref{eqt.densityQCD}$ for all QCD axions with $f_a<M_{pl}$. Taking the topological defect decay and other corrections into consideration,
\begin{align}
	\Omega_a h^2\approx\left\{
	\begin{matrix}
	&2\times 10^4\left(\frac{f_a}{10^{16}\text{GeV}}\right)^{7/6}(\pi^2/3)F_{\text{anh.}}(\pi/\sqrt{3})(1+\alpha_{\text{dec}}) & \text{PQ unbroken during inflation}\\
	&2\times 10^4\left(\frac{f_a}{10^{16}\text{GeV}}\right)^{7/6}\left(\theta_{i}^2+(\frac{H_I}{2\pi f_a})^2\right)F_{\text{anh.}}\left(\sqrt{\theta_{i}^2+(\frac{H_I}{2\pi f_a})^2}\right) & \text{PQ broken during inflation}
	\end{matrix}
\right. ~,
\end{align}
where the correction factor $F_{\text{anh.}}(x)=\left[1-\ln\left(1-x^2/\pi^2\right)\right]^{7/6}$, for the violation of the $x\ll 1$ assumption. In Fig.~\ref{fig.qcdadm}, we show the relic density of QCD axions, when PQ broke during inflation and QCD axions constitute cold DM today. While Fig.~\ref{fig.qcdadmu} shows for the PQ unbroken scenario.
\subsection{Cosmological constrains}\label{sec.constrain}
\begin{figure*}[]
	\centering
	\includegraphics[width=0.5\textwidth]{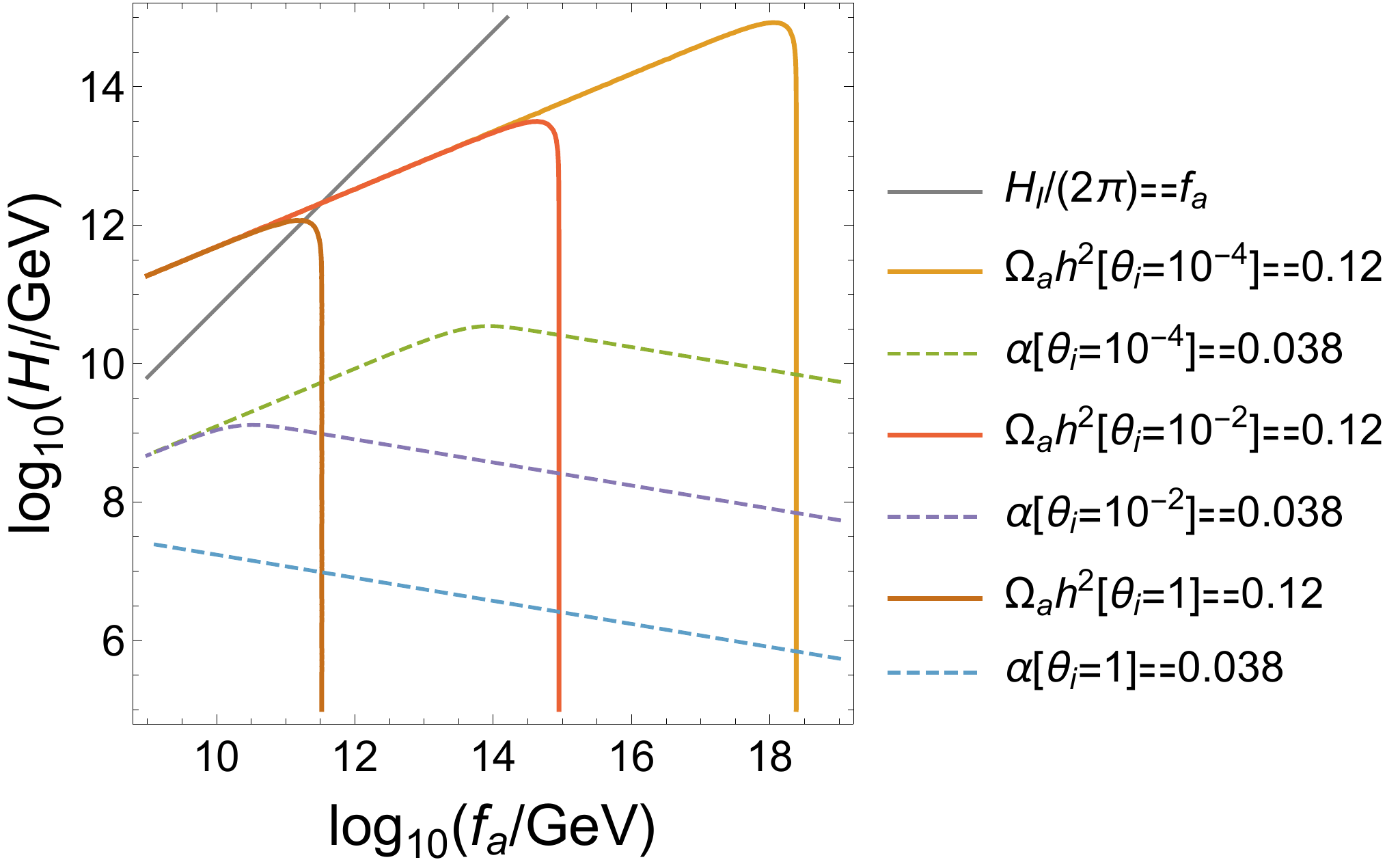}
	\caption{Constrains from dark matter abundance and isocurvature spectrum, for QCD axions produced in PQ broken scenario. Reproduced from \cite{Marsh:2015xka}. The solid lines present when axions constitute all the dark matter, and the region below the line is the parameter space allowed. 
The dashed lines present contribution of axion perturbations, constrained by isocurvature ratio $\alpha=A_{I,a}/A_s<0.038$, and the region below is allowed. The three groups of lines are for different initial misalignment angle $\theta_i$. The condition for PQ broken scenario corresponds to below the grey line. 
We can see that, for a high scale inflation, only small $\theta_i$ is possible.\ \black }
	\label{fig.QCDiso}
\end{figure*}

\begin{figure*}[t]
	\centering
	\includegraphics[width=0.55\textwidth]{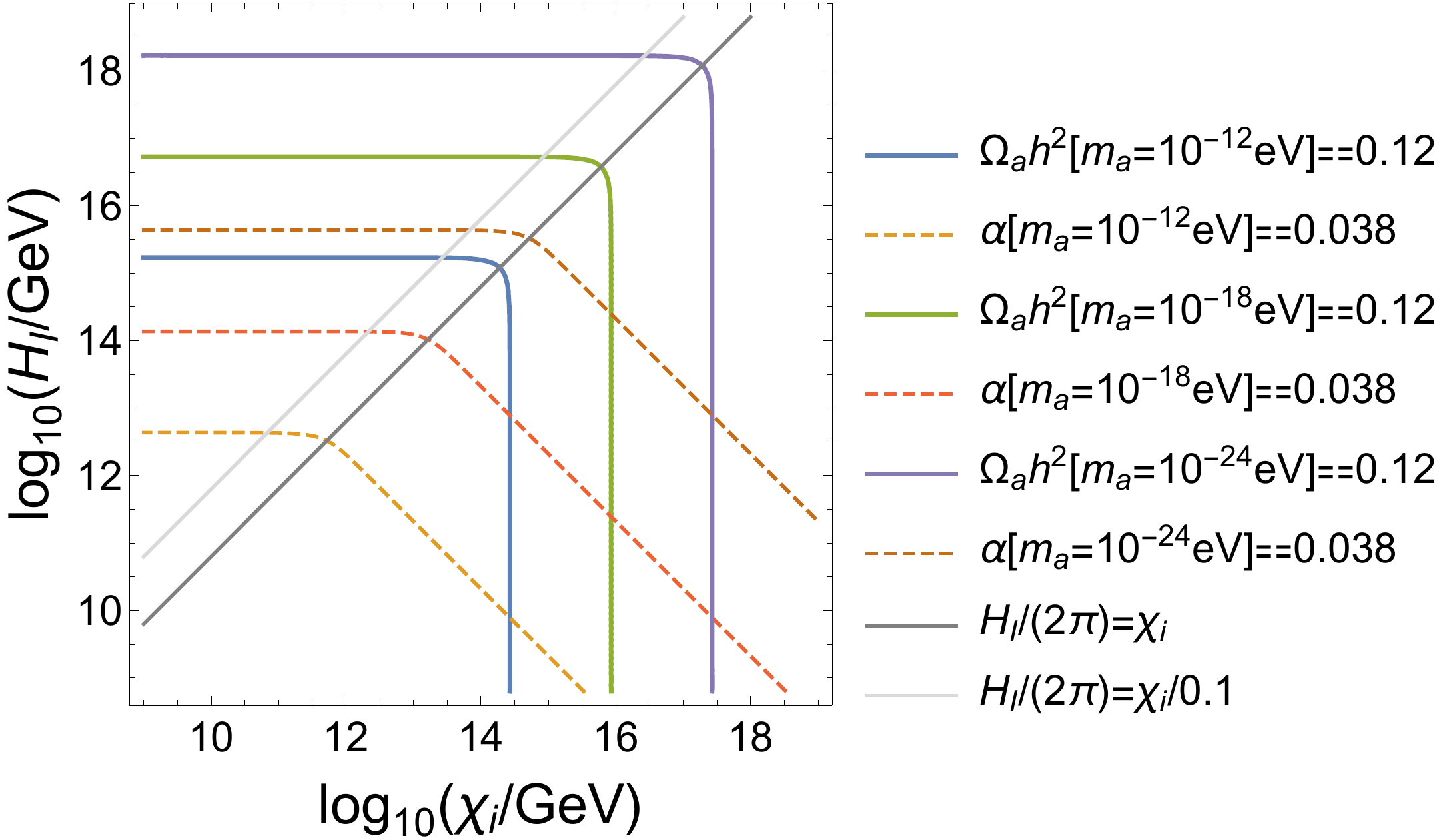}
	\caption{Constrains from dark matter abundance and isocurvature spectrum, for ALPs production in PQ broken scenario, under the assumption that ALPs began oscillation during RD epoch. The solid lines present when axion constitute all the dark matter, and the region below the line is the parameter space allowed. The dashed lines present when axion perturbations contribute all the isocurvature perturbation, and the region below is allowed. The three groups of lines are for ALPs with different masses $m_a$. PQ broken scenario requires $H_I/(2\pi)<f_a=\chi_i/\theta_i$, which is showed by gray lines for different initial misalignment angle $\theta_i$. We can see that, lighter ALPs allow higher scale inflation.}
	\label{fig.ALPmaiso}
\end{figure*}

The axion fluctuations are converted to isocurvature perturbations, as $S_{ij}=3(\zeta_i-\zeta_j)$. For axion dark matter\cite{Fox:2004kb} 
\begin{align}
	S_a\simeq \frac{\Omega_a}{\Omega_{d}}\frac{\delta\rho_a}{\rho_a}=\frac{\Omega_a}{\Omega_{d}}\frac{\delta(\theta^2)}{\langle \theta^2\rangle}=\left(\frac{\Omega_a}{\Omega_{d}}\right)\frac{(\theta_{i}+\delta\theta)^2-\langle\theta^2\rangle}{\langle \theta^2\rangle}
=\left(\frac{\Omega_a}{\Omega_{d}}\right)\frac{-\sigma_a^2+2\theta_{i}\delta\theta+(\delta\theta)^2}{\theta_{i}^2+\sigma_a^2} ~,
\label{eqt.sa}
\end{align}
where $\Omega_a/\Omega_d$ is the axion fraction of cold dark matter, the expectation value $\langle\theta^2 \rangle=(\theta_{i}^2+\sigma_a^2)$, $\theta_{i}$ is the initial misalignment angle at PQ broken, and $\sigma_a=H_I/(2\pi f_a)$. The isocurvature power spectrum contributed by axions have an amplitude of
\begin{align}
	A_{I,a}=\left(\frac{\Omega_a}{\Omega_{d}}\right)^2\frac{2\sigma_a^2(2\theta_{i}^2+\sigma_a^2)}{(\theta_{i}^2+\sigma_a^2)^2} ~,
\end{align}
while the isocurvature mode is constrained by
\begin{align}
	\beta_{\text{iso},a}\simeq\frac{A_{I,a}}{A_s}<0.038 ~,
\end{align}
where the scalar power spectrum amplitude $A_s=2.2\times 10^{-9}$, and we have taken isocurvature perturbations as uncorrelated and use the data in \cite{Ade:2015lrj}.
Combined with another constrain from dark matter production,
\begin{align}
	\Omega_a h^2\leq\Omega_d h^2=0.12 ~,
\end{align}
	we can show the constrains in one diagram. For QCD axion case, we get Fig.~\ref{fig.QCDiso} for different initial misalignment angle $\theta_i$. It can be seen that, the parameter space for high-scale inflation is severely constrained, and only small $\theta_i$ is allowed. 
\\ \indent
For ALPs from high-dimensional theories, we show cases with different axion masses $m_a$ in Fig.~\ref{fig.ALPmaiso}. Constrains for parameters of this model are much relaxed.
\section{Fluctuations during inflation}\label{sec.fluc}
During inflation, light fields fluctuate due to quantum effects, with an amplitude $H_I/2\pi$ equals to the Gibbons-Hawking temperature \cite{Gibbons:1977mu}. To study the collider signal recorded by axion fluctuations, first we need to deal with the axion interaction and the fluctuation modes.
\subsection{Interaction}
In this paper we will focus on the interaction between an axion and a massive vector field $Z_\mu$, as in \eqref{eqt.interactZ}. Write the action
\begin{align}
	S=\int d^4x\sqrt{-g}\left[-\frac{1}{4}(F_{\mu\nu})^2-\frac{1}{2}m^2_Z Z_{\mu}^2-\frac{1}{4}c_0\theta F_{\mu\nu}\tilde{F}^{\mu\nu}+...\right] ~,
\end{align}
where $\theta$ is the axion field $\chi$ divided by axion decay constant $f_a$, the coefficient $c_0$ is a constant usually of $\mathcal{O}(1)$ according to \eqref{eqt.cog}. In de-Sitter space the metric is $g_{\mu\nu}=a^2(-1,1,1,1)$, where $a$ is the time dependent scale factor. For convenience we rewrite the above action as
\begin{align}
	S=\int d^4x\left[-\frac{1}{4}F_{\mu\nu}F^{\mu\nu}-\frac{1}{2}a^2 m^2_Z Z_{\mu}Z^{\mu}-\frac{1}{4}c_0\theta F_{\mu\nu}F_{\rho\sigma}\epsilon^{\mu\nu\rho\sigma}\right] ~,
\end{align}
where $\epsilon^{\mu\nu\rho\sigma}$ is Levi-Cevita symbol, and the index transform as if in Minkowski spacetime with the metric $\eta_{\mu\nu}=(-1,1,1,1)$.
Then we can get the equation of motion for $Z_\mu$
\begin{align}
	(-\partial^2+a^2m^2_Z)Z^{\sigma}+\partial_\rho\partial^\sigma Z^\rho=2 c_0(\partial_\rho \theta)\epsilon^{\mu\nu\rho\sigma}\partial_\mu Z_\nu ~.
\label{eqt.eom}
\end{align}
\indent
\subsection{Propagators}
\indent
Split the axion field into background and quantum fluctuations $\chi=\chi_0+\delta\chi$. During inflation, the momentum-space propagator for $\delta \theta=\delta\chi/f_a$ from $\tau$ to $\tau_f$ is
\begin{align}
	G^{a}(\mathbf{k},\tau)=G^{a +}(k;\tau,\tau_f)=\frac{1}{f_a^2}\frac{H^2}{2k^3}(1-i a k\tau)e^{i a k\tau}
\label{eqt.proG} ~,
\end{align}
where $\tau$ is the conformal time, $\tau_f$ is the time when axion fluctuation exits the horizon, which is usually taken to be $0$, and $a=\pm$ denotes different types of vertices for Schwinger-Keldysh time integration (see details in \cite{Chen:2017ryl}).\\
\indent
Massive vector fluctuations have three polarization modes in the unitary gauge. Quantization of $Z_\mu$ is
\begin{align}
	Z_{\mu,\mathbf{p}}(\tau)=\text{v}_\mu(\mathbf{p},\tau)\mathnormal{a}_\mathbf{p}+\text{v}^*_\mu(-\mathbf{p},\tau)a^\dagger_\mathbf{-p} ~,
\end{align}
where
\begin{align}
	\text{v}_\mu(\mathbf{p},\tau)=v^{\scriptscriptstyle(+)\displaystyle}_p(\tau)\epsilon^{\scriptscriptstyle{(+)}}_{\mu,\mathbf{p}}+v^{\scriptscriptstyle(-)}_p(\tau)\epsilon^{\scriptscriptstyle(-)}_{\mu,\mathbf{p}}+v^{\scriptscriptstyle(\parallel)}_p(\tau)\epsilon^{\scriptscriptstyle(\parallel)}_{\mu,\mathbf{p}} ~,
\end{align}
$v_p^{(\lambda)}$ are mode functions for three polarizations ($\lambda$=$\pm$ or $\parallel$), which can be solved from (\ref{eqt.eom}). $Z^0$ has no dynamic in unitary gauge\cite{Liu:2019fag}. Then the equation of motion for vector fluctuations $Z_\mu$ becomes
\begin{align}
	(\partial_\tau^2+p^2+m^2_Z a^2)\mathbf{Z}=2ic_0a\dot{\theta}_0\mathbf{p\times Z} ~,
\end{align}
where we do not ignore the axion background $\dot\chi_0=f_a\dot{\theta}_0$. The background evolution of $\chi$ is ``slow-rolling" if its energy density is subdominant during inflation, and we treat it as a constant in the following \footnote{$\dot{\theta}_0\simeq \frac{\Lambda_\chi^4}{f_a^2(3H+\Gamma)}$, where $\Lambda_\chi$ is the scale factor of axion potential $V(\theta)=\Lambda_\chi^4(1-\cos\theta)$, and $\Gamma$ is due to the backreaction of $Z$ production.}. Then we split the components of 
$\mathbf{Z}$ into $\mathbf{Z^\bot}+(\mathbf{Z\cdot}\mathbf{\epsilon_p}^{\scriptscriptstyle(\parallel)})\mathbf{\epsilon_p}^{\scriptscriptstyle(\parallel)}$, where the longitudinal polarization unit vector $\mathbf{\epsilon}^{\scriptscriptstyle(\parallel)}_{\mathbf{p}}=\mathbf{\hat{p}}=\mathbf{p}/p$. For transverse components, we chose circular polarizations $\mathbf{\epsilon}^{\scriptscriptstyle(\pm)}_{\mathbf{p}}$, which are defined as
\begin{align}
	\mathbf{\epsilon}_\mathbf{p}^{\scriptscriptstyle{(\pm)}}=\frac{1}{\sqrt{2(1-(\mathbf{\hat{n}\cdot\hat{p}})^2)}}[(\mathbf{\hat{n}}-(\mathbf{\hat{n}\cdot\hat{p}})\mathbf{\hat{p}})\pm i(\mathbf{\hat{p}\times\hat{n}})] ~,
\end{align}
where $\mathbf{\hat{n}}$ is a random unit vector different from $\mathbf{\hat{p}}$. We have
\begin{align}
	\epsilon_{i,\mathbf{p}}^{\scriptscriptstyle{(\pm)}}(\epsilon_{j,\mathbf{p}}^{\scriptscriptstyle{(\pm)}})^*=\frac{1}{2}\left(\delta_{ij}-\frac{p_ip_j}{p^2}\mp i \frac{p_k}{p} \epsilon^{ijk}
\right) ~,
\label{eqt.contractpolar}
\end{align}
independent of the choice of $\mathbf{\hat{n}}$.
The mode functions can be solved from
\begin{align}
	&v_p^{\scriptscriptstyle{(\pm)}\prime\prime}(\tau)+\left(p^2\mp \frac{2pc_0\dot{\theta}_0}{H\tau}+\frac{m^2_Z}{H^2\tau^2}
\right)v_p^{\scriptscriptstyle{(\pm)}}(\tau)=0 ~,\\
	&v_p^{\scriptscriptstyle{(\parallel)}\prime\prime}(\tau)+\left(p^2+\frac{m^2_Z}{H^2\tau^2}
\right)v_p^{\scriptscriptstyle{(\parallel)}}(\tau)=0 ~,
\end{align}
as
\begin{align}
	v^{\scriptscriptstyle{(\pm)}}_p(\tau)&=\frac{1}{\sqrt{2p}}2^{\mp ic}e^{\mp\frac{\pi}{2}c}W(\pm ic,i\mu,2ip\tau) ~,\\
	v^{\scriptscriptstyle{(\parallel)}}_p(\tau)&=\frac{1}{\sqrt{2p}}W(0,i\mu,2ip\tau)=
i\frac{\sqrt{\pi}}{2}e^{-i\pi/4}(-\tau)^{1/2}e^{-\pi\mu/2}H^{(1)}_{i\mu}(-p\tau) ~,
\end{align}
under Bunch-Davies conditions, where $W$ is the Whittaker function, and two dimensionless parameters $\mu=\sqrt{(m_Z/H)^2-1/4}$, $c=c_0\dot{\theta}_0/H$. If we take late time limit $(-p\tau)\rightarrow 0$,
\begin{align}
	v^{\scriptscriptstyle{(\pm)}}_p(\tau)\simeq e^{-i\pi/4}2^{\mp ic}e^{\mp\pi c/2}(-\tau)^{1/2}&\left[\frac{\Gamma(-2i\mu)}{\Gamma(1/2\mp ic-i\mu)}e^{\pi\mu/2}(-2p\tau)^{i\mu}+(\mu\rightarrow -\mu)
\right] ~,\\
	v^{\scriptscriptstyle{(\parallel)}}_p(\tau)\simeq e^{-i\pi/4}(-\tau)^{1/2}&\left[\frac{\Gamma(-2i\mu)}{\Gamma(1/2-i\mu)}e^{\pi\mu/2}(-2p\tau)^{i\mu}+(\mu\rightarrow -\mu)
\right] ~,
\label{eqt.vlimit}
\end{align}
for real $\mu$ and $c$.
Now we can write propagators for $Z$
\begin{align}
	D^{\scriptscriptstyle{-+}}_{i'j}(\mathbf{p},\tau',\tau)=\sum_{\scriptscriptstyle{\lambda=\pm,\parallel}}\mathbf{\epsilon}^{(\lambda)}_{i',\mathbf{p}}v^{(\lambda)}_p(\tau')
\left[\mathbf{\epsilon}^{(\lambda)}_{j,\mathbf{p}}v^{(\lambda)}_p(\tau)\right]^*&=D^{\scriptscriptstyle{>}}_{i'j}(\mathbf{p},\tau',\tau) ~,\\
	D^{\scriptscriptstyle{+-}}_{i'j}(\mathbf{p},\tau',\tau)=\sum_{\scriptscriptstyle{\lambda=\pm,\parallel}}\left[\mathbf{\epsilon}^{(\lambda)}_{i',-\mathbf{p}}v^{(\lambda)}_p(\tau')
\right]^*\mathbf{\epsilon}^{(\lambda)}_{j,-\mathbf{p}}v^{(\lambda)}_p(\tau)&=[D^{\scriptscriptstyle{-+}}_{i'j}(\mathbf{p},\tau',\tau)]^*_{\mathbf{p}\rightarrow -\mathbf{p}} ~,
\end{align}
and
\begin{align}\nonumber
	D^{\scriptscriptstyle{++}}_{i'j}(\mathbf{p},\tau',\tau)&=\mathrm{\theta}(\tau'-\tau)D^{\scriptscriptstyle{-+}}_{i'j}(\mathbf{p},\tau',\tau)+\mathrm{\theta}(\tau-\tau')D^{\scriptscriptstyle{+-}}_{i'j}(\mathbf{p},\tau',\tau) ~,\\
	D^{\scriptscriptstyle{--}}_{i'j}(\mathbf{p},\tau',\tau)&=\mathrm{\theta}(\tau'-\tau)D^{\scriptscriptstyle{+-}}_{i'j}(\mathbf{p},\tau',\tau)+\mathrm{\theta}(\tau-\tau')D^{\scriptscriptstyle{-+}}_{i'j}(\mathbf{p},\tau',\tau)
=[D^{\scriptscriptstyle{++}}_{i'j}(\mathbf{p},\tau',\tau)]^*_{\mathbf{p}\rightarrow -\mathbf{p}}
~.
\end{align}
\indent
Making use of the polarization relations (\ref{eqt.contractpolar}), we have
\begin{align}\nonumber
	D^{\scriptscriptstyle{-+}}_{i'j}(\mathbf{p},\tau',\tau)=\frac{1}{2}(\delta_{i'j}-p_{i'}p_j/p^2&-i\epsilon^{i'jk}p_k/p)u^{\scriptscriptstyle(+)}_p(\tau',\tau)&\\ \nonumber
+\frac{1}{2}(\delta_{i'j}-p_{i'}p_j/p^2&+i\epsilon^{i'jk}p_k/p)u^{\scriptscriptstyle(-)}_p(\tau',\tau)&\\
+ p_{i'}p_j/p^2&\ u^{\scriptscriptstyle(\parallel)}_p(\tau',\tau) ~,&
\end{align}
where $u^{(\lambda)}_p(\tau',\tau)=v^{(\lambda)}_p(\tau')(v^{(\lambda)}_p(\tau))^*$. For convenience of contracting indices, we separate propagator into three parts with different dependence of the momentum direction
\begin{align}
	D^{ba}_{i'j}(\mathbf{p},\tau',\tau)=A^{ba}(p,\tau',\tau)\delta_{i'j}+B^{ba}(p,\tau',\tau)p_{i'}p_j+C^{ba}(p,\tau',\tau)\epsilon^{i'jk}p_k ~,
\end{align}
where
\begin{align}
	A^{\scriptscriptstyle{-+}}(p,\tau',\tau)&=\frac{1}{2}[u^{\scriptscriptstyle(+)}_p(\tau',\tau)+u^{\scriptscriptstyle(-)}_p(\tau',\tau)]=A^{\scriptscriptstyle{>}}(p,\tau',\tau) ~,\\
	B^{\scriptscriptstyle{-+}}(p,\tau',\tau)&=\{u^{\scriptscriptstyle(\parallel)}_p(\tau',\tau)-\frac{1}{2}[u^{\scriptscriptstyle(+)}_p(\tau',\tau)+u^{\scriptscriptstyle(-)}_p(\tau',\tau)]\}\frac{1}{p^2} ~,\\
	C^{\scriptscriptstyle{-+}}(p,\tau',\tau)&=-\frac{i}{2}[u^{\scriptscriptstyle(+)}_p(\tau',\tau)-u^{\scriptscriptstyle(-)}_p(\tau',\tau)]\frac{1}{p} ~,
\end{align}
only depend on the momentum amplitude, and correspondingly,
\begin{align}
	A^{\scriptscriptstyle{+-}}(p,\tau',\tau)&=[A^{\scriptscriptstyle{-+}}(p,\tau',\tau)]^*=A^{\scriptscriptstyle{<}}(p,\tau',\tau) ~,\\
	B^{\scriptscriptstyle{+-}}(p,\tau',\tau)&=[B^{\scriptscriptstyle{-+}}(p,\tau',\tau)]^* ~,\\
	C^{\scriptscriptstyle{+-}}(p,\tau',\tau)&=-[C^{\scriptscriptstyle{-+}}(p,\tau',\tau)]^* ~.
\end{align}
Note terms $B$ and $C$ survive only when $c\neq 0$.\\
\indent
Now we have the complete form of all propagators in S-K formalism. Note that for propagators with the same sign for $a$ and $b$, $\delta(\tau-\tau')$ term may appear in time integration if there is second time derivative. This can be written as\cite{Chen:2017ryl}
\begin{align} \nonumber
	\partial_{\tau'}\partial_{\tau}A^{\scriptscriptstyle{++}}(p,\tau',\tau)&=[\partial_{\tau'}\partial_{\tau}A^{\scriptscriptstyle{>}}(p,\tau',\tau)]\theta(\tau-\tau')+[\partial_{\tau'}\partial_{\tau}A^{\scriptscriptstyle{<}}(p,\tau',\tau)]\theta(\tau'-\tau)\\
&+\partial_\tau[A^{\scriptscriptstyle{>}}(p,\tau',\tau)-A^{\scriptscriptstyle{<}}(p,\tau',\tau)]\delta(\tau'-\tau) ~,
\end{align}
and similarly for $B$ and $C$. This extra delta term truly has physical meaning, so need to be included in calculation. As $\partial_{\tau}[u^{\scriptscriptstyle{(\lambda)}}_p(\tau',\tau)-u^{\scriptscriptstyle{(\lambda)}}_p(\tau',\tau)^*]|_{\tau'=\tau}=i$ for $\lambda=(\pm,\parallel)$,
\begin{align}
	\partial_\tau[A^{\scriptscriptstyle{>}}(p,\tau',\tau)-A^{\scriptscriptstyle{<}}(p,\tau',\tau)]=i ~,
\end{align}
while
\begin{align}
	\partial_\tau[B^{\scriptscriptstyle{>}}(p,\tau',\tau)-B^{\scriptscriptstyle{<}}(p,\tau',\tau)]=0 ~,\\
	\partial_\tau[C^{\scriptscriptstyle{>}}(p,\tau',\tau)-C^{\scriptscriptstyle{<}}(p,\tau',\tau)]=0 ~.
\end{align}
\section{Non-Gaussianity from axion relic perturbation}\label{sec.nonGau} 
To simplify the calculation, we can make a substitution for heavy Z using \textit{partial effective field theory}\cite{Iyer:2017qzw}
\begin{align}
	Z^\delta
\rightarrow 
	\frac{2c_0}{a^2m^2_Z}\epsilon^{\alpha\beta\gamma\delta}(\partial_\gamma\theta)\partial_\alpha Z_\beta ~,
\end{align}
from \eqref{eqt.eom}. To preserve leading order non-local effect, we only substitue one of the two vector fields in the interaction
\begin{align}\nonumber
	-\frac{1}{4}c_0\theta F_{\mu\nu}F_{\rho\sigma}\epsilon^{\mu\nu\rho\sigma}=&-c_0\epsilon^{\mu\nu\rho\sigma}\theta \partial_{\mu}Z_\nu\partial_{\rho}Z_\sigma\\
\rightarrow&
	-c_0\epsilon^{\mu\nu\rho\sigma}\theta \partial_{\mu}Z_\nu\partial_{\rho}\left(\eta_{\sigma\delta}\frac{2c_0}{a^2m^2_Z}\epsilon^{\alpha\beta\gamma\delta}(\partial_\gamma\theta)\partial_\alpha Z_\beta\right)\\
	=&\frac{2c^2_0}{a^2m^2_Z}\eta_{\sigma\delta}\epsilon^{\mu\nu\rho\sigma}\epsilon^{\alpha\beta\gamma\delta}(\partial_\rho \theta \partial_\mu Z_\nu)(\partial_\gamma \theta\partial_\alpha Z_\beta) ~,
\end{align}
where at the last step, we did integration by part and used properties of the Levi-Cevita symbol. This can be treated as an effective 4-vertex, then we draw the diagram for leading order contribution to $\langle\delta\theta\delta\theta\delta\theta\rangle$,
\begin{figure*}[h]
	\centering
	\includegraphics[width=0.4\textwidth]{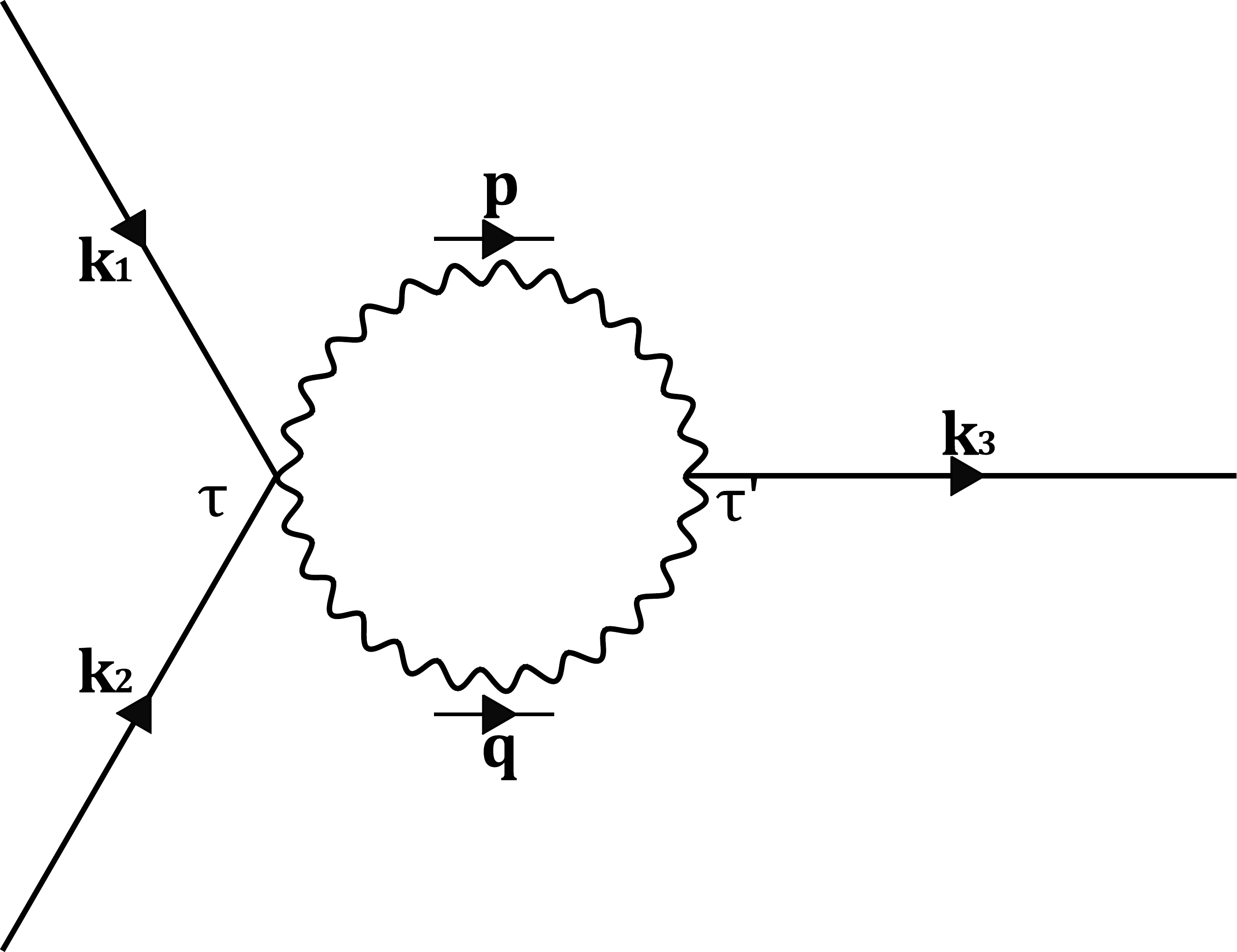}
\end{figure*}
\\
and
\begin{align}\nonumber
\langle\delta\theta(\mathbf{k_1})\delta\theta(\mathbf{k_2})\delta\theta(\mathbf{k_3})\rangle'
	=&\ \sum_{ab=\pm}ab\int d\tau'\int d\tau\ 2(-ic_0)i\frac{2c^2_0}{m^2_Z}(H\tau)^2 G^b(\mathbf{k_3},\tau')\left[\text{D}_\rho G^a(\mathbf{k_1},\tau)\text{D}_{\rho_1} G^a(\mathbf{k_2},\tau)+\right.\\
\left.+(\mathbf{k_1}\leftrightarrow\mathbf{k_2})
\right]
&\int\frac{\mathbf{d}^3\mathbf{p}}{(2\pi)^3}\left[\epsilon^{\alpha'\beta'\gamma'\delta'}\epsilon^{\mu\nu\rho\sigma}\epsilon^{\mu_1\nu_1\rho_1\sigma_1}\eta_{\sigma\sigma_1}\text{D}_{\alpha'}\text{D}_\mu D^{\ b a}_{\beta'\nu}(\mathbf{p},\tau',\tau)\text{D}_{\gamma'}\text{D}_{\mu_1} D^{\ b a}_{\delta'\nu_1}(\mathbf{q},\tau',\tau)
\right] ~,
\label{eqt.diagram}
\end{align}
where G denotes the propagators for $\theta$ fluctuation $\delta\theta=\delta \chi/f$, which we treat as massless scalar field here, and D denotes propagators for vector field. $D_\mu$ here is time derivative when it denotes the time component, but relates to the propagating momentum when in momentum space. For example,
\begin{align}\nonumber
	\text{D}_\rho G^a (\mathbf{k_1},\tau)&=\left(\begin{matrix}\partial_{\tau}\\ i\mathbf{k_1}\end{matrix}\right) G^a(k_1,\tau) ~,\\
	\text{D}_{\alpha'}\text{D}_\mu D^{\ b a}_{\beta'\nu} (\mathbf{p},\tau',\tau)&=\left(\begin{matrix}\partial_{\tau'}\\ i\mathbf{p}\end{matrix}\right)\left(\begin{matrix}\partial_{\tau}\\ -i\mathbf{p}\end{matrix}\right)D^{\ b a}_{\beta'\nu}(\mathbf{p},\tau',\tau) ~.
\end{align}
\subsection{Integral}
\indent
With these we can go back to $\langle\delta\theta_{\mathbf{k_1}}\delta\theta_{\mathbf{k_2}}\delta\theta_{\mathbf{k_3}}\rangle'$, as presented in (\ref{eqt.diagram}). First of all, we contract all the indices in the following part
\begin{align}\nonumber
	\epsilon^{\alpha'\beta'\gamma'\delta'}\epsilon^{\mu\nu\rho\sigma}\epsilon^{\mu_1\nu_1\rho_1\sigma_1}\eta_{\sigma\sigma_1}\text{D}_{\alpha'}\text{D}_\mu D^{\ b a}_{\beta'\nu}(\mathbf{p},\tau',\tau)\text{D}_{\gamma'}\text{D}_{\mu_1} D^{\ b a}_{\delta'\nu_1}(\mathbf{q},\tau',\tau)
	\left[\text{D}_\rho G^a(\mathbf{k}_\mathbf{1},\tau)\text{D}_{\rho_1} G^a(\mathbf{k}_\mathbf{2},\tau)\right.\\ \left.
+(\mathbf{k}_\mathbf{1}\leftrightarrow\mathbf{k}_\mathbf{2})\right] ~,
\end{align}
the whole result is shown in the Appendix~\ref{sec.appendixA}. Then we can write (\ref{eqt.diagram}) as
\begin{align}
\nonumber
	 \frac{4c^3_0H^2}{m^2_Z}\int\frac{\mathbf{d}^3\mathbf{p}}{(2\pi)^3}\ \sum_{ab=\pm}ab\int^{0}_{-\infty} d\tau'\int^{0}_{-\infty}  d\tau\ \tau^2 G^b(\mathbf{k_3},\tau')[& I^{\ ba}_{AA}+I^{\ ba}_{AB}+I^{\ ba}_{BB}\\
+& I^{\ ba}_{BC}+I^{\ ba}_{CC}+I^{\ ba}_{AC}](\mathbf{k_1},\mathbf{k_2},\mathbf{p};\tau',\tau) ~,
\label{eqt.integral}
\end{align}
we can always write
\begin{align}
	G^b(\mathbf{k_3},\tau')[I^{\ ba}_{AA}+I^{\ ba}_{AB}+I^{\ ba}_{BB}+I^{\ ba}_{BC}+I^{\ ba}_{CC}+I^{\ ba}_{AC}](\mathbf{k_1},\mathbf{k_2},\mathbf{p};\tau',\tau) ~,
\label{eqt.integrand}
\end{align}
to be
\begin{align}
	G^b(\mathbf{k_3},\tau')[G^a(\mathbf{k_1},\tau)G^a(\mathbf{k_2},\tau)&I^{ba}_{0}(\mathbf{k_1},\mathbf{k_2},\mathbf{p},\mathbf{q};\tau',\tau)\\ 
\label{eqt.I1}
+\partial_\tau G^a(\mathbf{k_1},\tau)G^a(\mathbf{k_2},\tau)&I^{ba}_{1}(\mathbf{k_1},\mathbf{k_2},\mathbf{p},\mathbf{q};\tau',\tau)\\
\label{eqt.I2}
+\partial_\tau G^a(\mathbf{k_1},\tau)\partial_\tau G^a(\mathbf{k_2},\tau)&I^{ba}_{2}(\mathbf{k_1},\mathbf{k_2},\mathbf{p},\mathbf{q};\tau',\tau)+(\mathbf{k_1}\leftrightarrow \mathbf{k_2})] ~,
\end{align}
so the S-K time integration of the first line can be written as
\begin{align}
	&\sum_{ab=\pm}ab\int^{0}_{-\infty} d\tau\int^{0}_{-\infty}  d\tau'\ \tau^2 G^b(\mathbf{k_3},\tau')G^a(\mathbf{k_1},\tau)G^a(\mathbf{k_2},\tau)I^{ba}_{0}(\mathbf{k_1},\mathbf{k_2},\mathbf{p},\mathbf{q};\tau',\tau)
\\ \label{eqt.Imp}
	=&\left[-\int^{0}_{-\infty} d\tau\int^{0}_{-\infty}  d\tau'\ \tau^2 G^-(\mathbf{k_3},\tau')G^+(\mathbf{k_1},\tau)G^+(\mathbf{k_2},\tau)I^{-+}_{0}(\mathbf{k_1},\mathbf{k_2},\mathbf{p},\mathbf{q};\tau',\tau)\right.
\\ \label{eqt.Ipp1}
	&+\int^{0}_{-\infty} d\tau\int^{0}_{-\infty}  d\tau'\ \tau^2 G^+(\mathbf{k_3},\tau')G^+(\mathbf{k_1},\tau)G^+(\mathbf{k_2},\tau)I^{+-}_{0}(\mathbf{k_1},\mathbf{k_2},\mathbf{p},\mathbf{q};\tau',\tau) 
\\ \label{eqt.Ipp2}
	&+\int^{0}_{-\infty} d\tau\int^{0}_{\tau}  d\tau'\ \tau^2 G^+(\mathbf{k_3},\tau')G^+(\mathbf{k_1},\tau)G^+(\mathbf{k_2},\tau)[I^{-+}_{0}-I^{+-}_{0}](\mathbf{k_1},\mathbf{k_2},\mathbf{p},\mathbf{q};\tau',\tau)
\\ \label{eqt.Idelta}
	&\left.+\int^{0}_{-\infty} d\tau\int^{0}_{-\infty}  d\tau'\ \tau^2 G^+(\mathbf{k_3},\tau')G^+(\mathbf{k_1},\tau)G^+(\mathbf{k_2},\tau)\delta(\tau'-\tau)U^{+}_{0}(\mathbf{k_1},\mathbf{k_2},\mathbf{p},\mathbf{q};\tau',\tau)\right] 
\\ \label{eqt.Iconj}
	+&c.c.\text{ with }(\mathbf{p}\rightarrow\mathbf{-p}) ~,
\end{align}
where \eqref{eqt.Idelta} appears when there is delta term coming from $\partial_{\tau'}\partial_{\tau}A^{\scriptscriptstyle{++}}$, and $U^{+}_{0}$ is the remaining part after delta function subtracted. The other parts \eqref{eqt.I1}\eqref{eqt.I2} can be also written in the same way.
\\
\indent
The full result is very tedious, thus we may take some limits or approximations as following,
\begin{center}
	\includegraphics[width=0.4\textwidth]{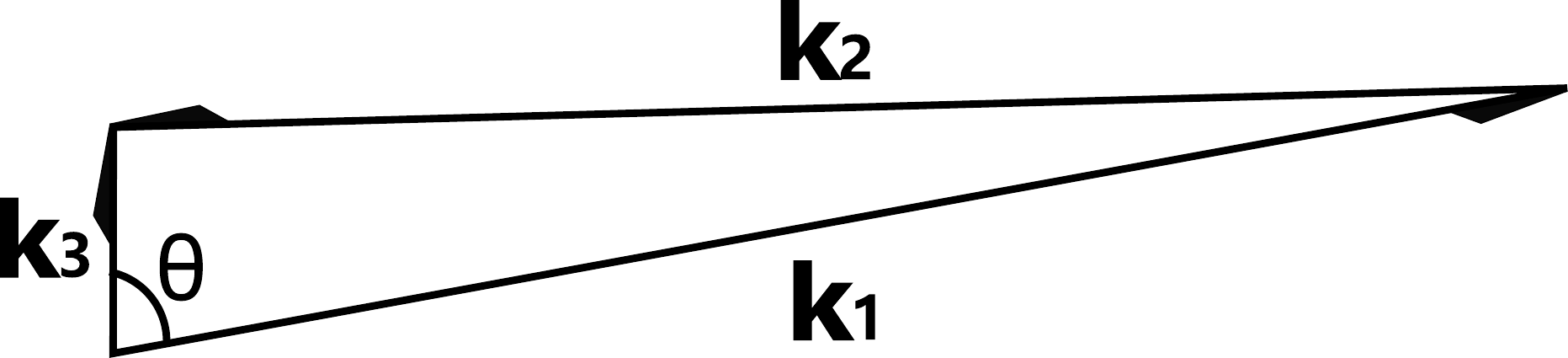}
\hspace*{70pt}
	\includegraphics[width=0.1\textwidth]{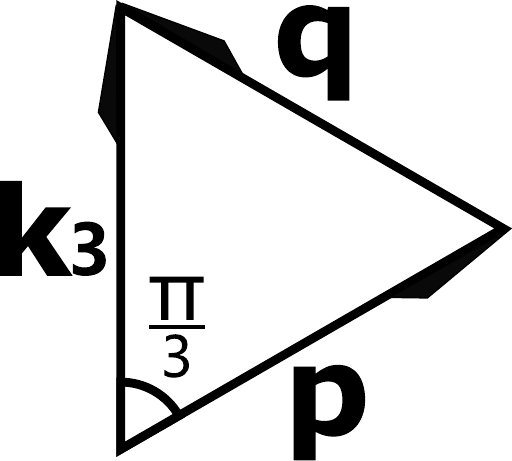}
\end{center}
\begin{itemize}
	\item Squeezed limit. For a cosmological collider, we are interested in the clock signal as $k_1\simeq k_2\gg k_3$.
\end{itemize}
\begin{align}
	\mathbf{k_3}&=(0,0,k_3)~,\ \mathbf{k_1}=(k_1\sin \theta,0,k_1\cos \theta)~,\ \mathbf{k_2}=-\mathbf{k_1}+\mathbf{k_3}
\label{eqt.momentumk}
\end{align}
\begin{itemize}
	\item Loop integral\cite{Chen:2018xck}\cite{Hook:2019zxa}. 
Instead of a rigorous calculation with renormalization procedure, we consider only a cutoff of internal wavelength $|\Lambda_q|\simeq k_3$. As an estimation, we can simply take $p\simeq q\simeq k_3$ for mode functions, so we can write
\end{itemize}
\begin{align}
	\mathbf{p}&=k_3\left(\frac{\sqrt{3}}{2}\cos \varphi_1,\frac{\sqrt{3}}{2}\sin \varphi_1,\frac{1}{2}\right) ~,\ 
	\mathbf{q}=k_3\left(-\frac{\sqrt{3}}{2}\cos \varphi_1,-\frac{\sqrt{3}}{2}\sin \varphi_1,\frac{1}{2}\right) ~,
\label{eqt.momentump}
\end{align}
\indent
the momentum integration becomes \(\int \frac{\mathbf{d}^3p}{(2\pi)^3}=\frac{k^3_3}{(2\pi)^3}\int^{2\pi}_0 d\varphi_1\), with only one angle $\phi_1$ to be integrated, which describes how far the plane of internal momenta rotates from the plane of external momenta.
\begin{itemize}
	\item Late time limit\cite{Chen:2016hrz}\cite{Chen:2018xck}. 
For vertex $\tau$ connecting to hard external line, late time limit $\tau\rightarrow 0$ can be a good approximation for  mode functions of internal fields. For vertex $\tau'$, approximation is good only for $0<p(-\tau')<\sqrt{\mu}$, while the resonance takes effect at $p(-\tau')\sim m_Z/H$.
	\item Large mass limit. 
We are considering heavy vector field with $m_Z\gtrsim\mathcal{O}(H)$, so $\mu\gtrsim\mathcal{O}(0.1)$ \footnote{This has been adopted when taking EFT method for interaction.}. We may take large $\mu$ limit to simplify $\mu$ functions and approximate the behavior of signal amplitude as the mass grows \footnote{For functions like $\Gamma[i\mu]$ and $\Gamma[i(c\pm\mu)]$.}. 
	\item Constrain for c. 
There is upper bound for $c$, e.g. if we wish the loop expansion brought by SM interaction not to diverge, which is amplified by factors like $e^{\pi c}$. With the help of EFT method, we can make the approximation\cite{Liu:2019fag}
\end{itemize}
\begin{equation}
	\raisebox{-0.45\totalheight}{\includegraphics[width=0.25\textwidth]{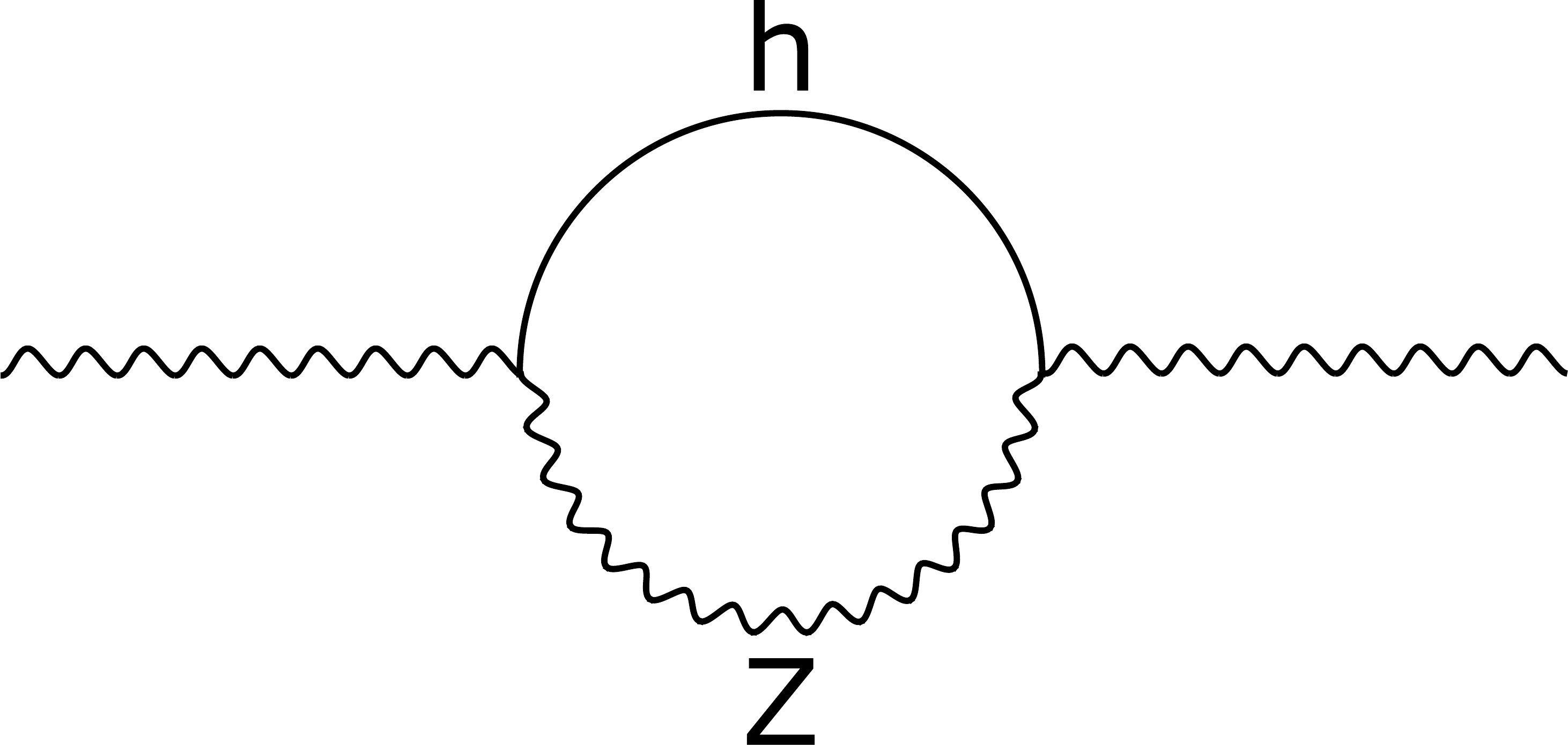}}\hspace*{10pt}\sim
\hspace*{10pt} \frac{g^2}{4\pi^2}\frac{m_Z^2}{m_h^2}e^{4\pi c}\times
	(\raisebox{+1\totalheight}{\includegraphics[width=0.15\textwidth]{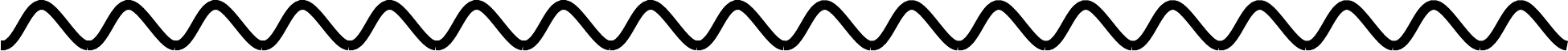}}) ~,
\end{equation}
\indent
to get $c\lesssim \frac{1}{2\pi}\log (\frac{4\pi m_h}{g m_Z})\sim \mathcal{O}(1)$ \footnote{The interaction $c_0\theta F\tilde{F}$ do not put extra constrain on c.}.
\subsection{Simplification}
\paragraph{Late time limit}
The propagators at late time limit $(-p\tau), (-p\tau')\rightarrow 0$ can be written from \eqref{eqt.vlimit}, for each polarization direction
\begin{align}\nonumber
	u^{\scriptscriptstyle(+)}_p(\tau',\tau) & =v^{\scriptscriptstyle(+)}_p(\tau')(v^{\scriptscriptstyle(+)}_p(\tau))^* \\ \nonumber
	& \simeq (-\tau')^{1/2}(-\tau)^{1/2}e^{-\pi c}\left[\frac{\Gamma(-2i\mu)\Gamma(-2i\mu)}{\Gamma(1/2- ic-i\mu)\Gamma(1/2 + ic - i\mu)}2^{2i\mu}(-p\tau')^{i\mu}(-p\tau)^{i\mu}+\right. \\
&  \left.+\frac{\Gamma(-2i\mu)\Gamma(2i\mu)}{\Gamma(1/2- ic-i\mu)\Gamma(1/2 + ic + i\mu)}e^{\pi\mu}(-p\tau')^{i\mu}(-p\tau)^{-i\mu}+(\mu\rightarrow -\mu)\right] ~,
\end{align}
and
\begin{align}
	u^{\scriptscriptstyle(-)}_p(\tau',\tau) & =u^{\scriptscriptstyle(+)}_p(\tau',\tau)|_{c\rightarrow -c} ~,\\
	u^{\scriptscriptstyle(\parallel)}_p(\tau',\tau) & =u^{\scriptscriptstyle(+)}_p(\tau',\tau)|_{c\rightarrow 0} ~,
\end{align}
for real $\mu$ and $c$. So the whole integrand in \eqref{eqt.integral} can be expanded into series of $\tau$ and $\tau'$, as shown in Appendix~\ref{sec.appendixB}.
\paragraph{Squeezed limit}
In the squeezed limit $k_1\simeq k_2\gg k_3$ and internal momentum limit $p\simeq q\simeq k_3$, we can work out the order of every term with $\tau$ and $\tau'$ integrated. So we are able to pick out the leading terms from \eqref{eqt.integrand}, which are underlined in Appendix~\ref{sec.appendixA}
\begin{align}\nonumber
I^{\ ba}_{BC,(0)}=
	\{i  \partial_{\tau'} \partial_{\tau}B^{ba}(q,\tau',\tau) \partial_{\tau} C^{ba}(p,\tau',\tau)[(\mathbf{p\times q\cdot}i\mathbf{k_1})(\mathbf{p\times q\cdot}i\mathbf{k_2})\hspace*{90pt}& \\ \nonumber 
    -(\mathbf{p\cdot q}) (\mathbf{p}\cdot i\mathbf{k_1}) (\mathbf{q}\cdot i\mathbf{k_2})+ q^2(\mathbf{p}\cdot i\mathbf{k_1})(\mathbf{p}\cdot i\mathbf{k_2})]
+(\mathbf{p}\leftrightarrow\mathbf{q}) \}G^a(k_1,\tau)G^a(k_2,\tau)& \\
+(\mathbf{k_1}\leftrightarrow\mathbf{k_2})& ~,
\label{eqt.IBC0}
\\
\nonumber
I^{\ ba}_{CA,(0)}=
	\{i \partial_{\tau'}\partial_{\tau}C^{ba}(p,\tau',\tau)\partial_{\tau}A^{ba}(q,\tau',\tau)[-(\mathbf{p}\cdot i\mathbf{k_1}) (\mathbf{q}\cdot i\mathbf{k_2})  - (\mathbf{p\cdot q})(i \mathbf{k_1}\cdot i\mathbf{k_2})]+(\mathbf{p}\leftrightarrow\mathbf{q}) \}&\\ \nonumber
 G^a(k_1,\tau)G^a(k_2,\tau)& \\
+(\mathbf{k_1}\leftrightarrow\mathbf{k_2})& ~,
\label{eqt.ICA0}
\\
\nonumber
I^{\ ba}_{AC,(0)}=
	\{i  \partial_{\tau'}\partial_{\tau}A^{ba}(q,\tau',\tau) \partial_{\tau}C^{ba}(p,\tau',\tau)[ p^2 (i \mathbf{k_1}\cdot i\mathbf{k_2}) + (\mathbf{p}\cdot i\mathbf{k_1})(\mathbf{p}\cdot i\mathbf{k_2}) ]+(\mathbf{p}\leftrightarrow\mathbf{q}) \}&\\ \nonumber
 G^a(k_1,\tau)G^a(k_2,\tau)& \\
+(\mathbf{k_1}\leftrightarrow\mathbf{k_2})& ~,
\label{eqt.IAC0}
\\
\nonumber
I^{\ ba}_{CC,(0)}=
	\{i \partial_{\tau'}\partial_{\tau}C^{ba}(p,\tau',\tau) \partial_{\tau}C^{ba}(q,\tau',\tau)  (\mathbf{q}\cdot i\mathbf{k_1})   (\mathbf{p\times q\cdot}i\mathbf{k_2})+(\mathbf{p}\leftrightarrow\mathbf{q}) \} G^a(k_1,\tau)G^a(k_2,\tau)& \\
+(\mathbf{k_1}\leftrightarrow\mathbf{k_2})& ~.
\label{eqt.ICC0}
\end{align}
Similar procedure can also exclude \eqref{eqt.Ipp2} and \eqref{eqt.Idelta} from S-K time integral cause the contribution is always sub-leading compared to others.
\paragraph{Momenta approximation}
With momenta written as \eqref{eqt.momentumk}\eqref{eqt.momentump}, the integral \eqref{eqt.integral} can be separated into two steps. Take $I^{\ ba}_{BC,(0)}$ \eqref{eqt.IBC0} as an example, first do the time integral
\begin{align}
	k_1^2k_3^4\sum_{ab=\pm}ab\int^{0}_{-\infty} d\tau'\int^{0}_{-\infty}  d\tau\ \tau^2 G^b(k_3,\tau')\partial_{\tau} C^{ba}(p,\tau',\tau) \partial_{\tau'} \partial_{\tau}B^{ba}(q,\tau',\tau)G^a(k_1,\tau)G^a(k_2,\tau) ~,
\label{eqt.IBCtime}
\end{align}
where $q=p=k_3$. Then collect the angle in momentum space
\begin{align}\nonumber
	k_1^{-2}k_3^{-4}\{i[(\mathbf{p\times q\cdot}i\mathbf{k_1})(\mathbf{p\times q\cdot}i\mathbf{k_2})
    -(\mathbf{p\cdot q}) (\mathbf{p}\cdot i\mathbf{k_1}) (\mathbf{q}\cdot i\mathbf{k_2})+ q^2(\mathbf{p}\cdot i\mathbf{k_1})(\mathbf{p}\cdot i\mathbf{k_2})]
+(\mathbf{p}\leftrightarrow\mathbf{q})\}&\\ \nonumber
+(\mathbf{k_1}\leftrightarrow\mathbf{k_2})& 
\\
\simeq i\left[(3\sin^2\theta\sin^2\varphi_1)+\frac{1}{2}(\cos^2\theta-3\sin^2\theta\cos^2\varphi_1)+(\cos^2\theta+3\sin^2\theta\cos^2\varphi_1)+\mathcal{O}\left(\frac{k_3}{k_1}\right)\right] ~,
\label{eqt.IBCangle}
\end{align}
and apply
$	\frac{4c^3_0H^2}{m^2_Z}\int\frac{\mathbf{d}^3\mathbf{p}}{(2\pi)^3}
=\frac{4c^3_0H^2}{m^2_Z}\frac{k_3^3}{(2\pi)^3}\int^{2\pi}_0 d\varphi_1
$ to \eqref{eqt.IBCangle}. 
Note that the angle integral make the contribution of $I^{\ ba}_{CC,(0)}$ vanish. So finally we only need to deal with $I^{\ ba}_{BC,(0)}$, $I^{\ ba}_{CA,(0)}$, $I^{\ ba}_{AC,(0)}$ and each with S-K time integral \eqref{eqt.Imp}\eqref{eqt.Ipp1}\eqref{eqt.Iconj}. Details are shown in Appendix~\ref{sec.appendixB}.
\subsection{Result}
The final result can be represented as
\begin{align}
\nonumber
	&\langle\delta\theta(\mathbf{k_1})\delta\theta(\mathbf{k_2})\delta\theta(\mathbf{k_3})\rangle'\\
&=\sum_{s=\scriptscriptstyle{BC,CA,AC}}F^s(\cos\theta)\times \frac{H_I^6}{f_a^6 k_1^6 k_3^3} \left[ f^s_1(\mu ,c)(k_1/k_3)^{2 i \mu }+f^s_2(\mu ,c)+ f^s_3(\mu ,c)(k_1/k_3)^{-2 i \mu }\right]\\
&=\frac{H_I^6}{f_a^6 k_1^6} \left[F_1(\cos\theta;\mu ,c)(k_1/k_3)^{2 i \mu }+F_2(\cos\theta;\mu ,c)+ F_3(\cos\theta;\mu ,c)(k_1/k_3)^{-2 i \mu }\right]
\end{align}
where $f^s_{1,2,3}(\mu,c)$ are from time integral, and $F^s(\cos\theta)$ is from the integration of angle $\varphi_1$. We have
\begin{align}\nonumber
	F_1(\cos\theta;\mu,c)&=\frac{c_0^3H^2}{\pi^2m_Z^2}
[f_\theta(\mu,c)\cos^2\theta+f_c(\mu,c)] ~,\\
	F_3(\cos\theta;\mu,c)&=F_1(\cos\theta;\mu,c)^* ~,
\label{eqt.fc}
\end{align}
$f^s_{1,2,3}$ and so $f_{\theta,c}$ can be obtained but the functions are still very complicated, so we do not write them here but just show some plots later.
\paragraph{Large $\mu$ limit}
We can make some other approximation if we wish to get a brief expression of the functions for the convenience to learn the signal amplitude as mass get large. Making use of \footnote{Actually $|y|$ do not have to go to infinity, the approximation has the right order for $|y|\geq\mathcal{O}(0.1)$.}
\begin{align}
	\Gamma(iy)\xrightarrow{|y|\rightarrow\infty} \sqrt{2 \pi } |y|^{-1/2} e^{-\pi |y|/2 } e^{i [y  ( \ln | y | -1) -\text{sgn}(y)\pi/4 )]}
\end{align}
where $y$ is real, and some procedures shown in Appendix~\ref{sec.appendixBmu}. We can write approximate expressions for $f_{\theta}$ and $f_c$ to second order when taking limits
\begin{align}
	f_\theta(\mu,c)
	\nonumber
	\xrightarrow[\mu>c>0]{\text{large }\mu,(\mu-c)}&
e^{-\pi  (\mu -c)} (1-e^{-2 \pi  c}) 2^{-12+2 i \mu } \pi^{1/2}  \mu ^{-3/2} (2 \mu +i)^3 (2 \mu +3 i) (\mu +3 i)\times\\ \nonumber
	\times&\left\{8 e^{-i \pi /4} (3 \mu +i) \mu ^{2 i \mu } (\mu -c)^{i (c-\mu )} (c+\mu )^{-i (c+\mu )}\right.\\ \nonumber
	& \color{darkgray} \left.+ e^{-\pi  (\mu -c)} i \pi^{1/2} \mu^{1/2} (2 \mu +i) (\mu +i)\left[11 (1+e^{-2 \pi  c}) \mu ^{4 i \mu } (\mu -c)^{2 i (c-\mu)} (c+\mu )^{-2 i (c+\mu )}\right.\right.\\  \nonumber
	& \color{darkgray} \ \ \left.\left.-6 e^{-\pi c} \mu ^{2 i \mu } (\mu -c)^{i (c-\mu )} (c+\mu )^{-i (c+\mu )}\right]+\mathcal{O}\left(e^{-2\pi(\mu-c)}\right)
\right\} ~,\\
	\nonumber
	\color{darkgray} \xrightarrow[c>\mu>0]{\text{large }\mu,(c-\mu)}& 
\color{darkgray} e^{4 \pi  (c-\mu )}11 * 2^{-12+2 i \mu }\pi \mu^{-1} (2 \mu +i)^3 (2 \mu +3 i) (\mu +3 i)\times\\
\nonumber
	\color{darkgray} \times &\color{darkgray} \left[ (1+e^{-2 \pi  \mu })^{-1}2 e^{i \pi /4} (\pi  \mu)^{-1/2} \mu ^{2 i \mu } (c-\mu )^{i (c-\mu )} (c+\mu )^{-i (c+\mu )}\right.\\
\nonumber
	&\color{darkgray}+ (1-e^{-4 \pi  \mu }) i(\mu +i) (2 \mu +i) \mu ^{4 i \mu } (c-\mu )^{2 i (c-\mu )} (c+\mu )^{-2 i (c+\mu )}\\
	&\color{darkgray}\left.+(1-e^{-4 \pi  \mu })^{-1}(\mu -i) (2 \mu -i)+\mathcal{O}\left(e^{-2\pi(c-\mu)}\right)\right] ~,
\label{eqt.fthetaa}
\end{align}
and
\begin{align}
	f_c(\mu,c)
	\nonumber
	\xrightarrow[\mu>c>0]{\text{large }\mu,(\mu-c)}&
e^{-\pi  (\mu -c)} (1-e^{-2 \pi  c}) 2^{-12+2 i \mu } \pi^{1/2} \mu ^{-3/2}(2 \mu +i)^3 (2 \mu +3 i) 	(\mu +3 i)\times \\ \nonumber
	\times & \left\{8 e^{3 i \pi /4}(5 \mu -3 i) \mu ^{2 i \mu } (\mu -c)^{i (c-\mu )} (c+\mu )^{-i (c+\mu )} \right.\\ \nonumber
	& \color{darkgray} +e^{-\pi  (\mu -c)}3 i \pi^{1/2} \mu^{1/2} (2 \mu +i) (\mu +i)\left[5 \left(1+e^{-2 \pi  c}\right) \mu ^{4 i \mu } (\mu -c)^{2 i (c-\mu )} (c+\mu )^{-2 i (c+\mu )}\right. \\ \nonumber
	& \color{darkgray} \ \ \left.\left.+6 e^{-\pi c} \mu ^{2 i \mu } (\mu -c)^{i (c-\mu )} (c+\mu )^{-i (c+\mu )} \right]+\mathcal{O}\left(e^{-2\pi(\mu-c)}\right)\right\} ~,\\
	\nonumber
	\color{darkgray} \xrightarrow[c>\mu>0]{\text{large }\mu,(c-\mu)}& 
\color{darkgray}  e^{4 \pi  (c-\mu )}15* 2^{-12+2 i \mu } \pi {\mu }^{-1} (2 \mu +i)^3 (2 \mu +3 i) (\mu +3 i)\times\\ \nonumber
	\color{darkgray} \times &\color{darkgray}  \left[(1+e^{-2 \pi  \mu })^{-1}2 e^{i\pi/4} \pi^{-1/2}\mu^{-1/2} \mu ^{2 i \mu }  (c-\mu )^{i (c-\mu )} (c+\mu )^{-i (c+\mu )}\right.\\
\nonumber
	&\color{darkgray} + (1-e^{-4 \pi  \mu }) i (\mu +i) (2 \mu +i) \mu ^{4 i \mu } (c-\mu )^{2 i (c-\mu )} (c+\mu )^{-2 i (c+\mu )}\\
	&\color{darkgray} \left.+(1-e^{-4 \pi  \mu })^{-1}(\mu -i) (2 \mu -i) + \mathcal{O}\left(e^{-2\pi(c-\mu)}\right) \right] ~,
\label{eqt.fca}
\end{align}
for large $|\mu-c|>\mathcal{O}(0.1)$ and $\mu>\mathcal{O}(0.1)$, and the cutoff at $\mathcal{O}\left(e^{-2\pi(c-\mu)}\right)$ works well for $|\mu-c|>\mathcal{O}(0.1)$.
\begin{figure*}[t]
	\begin{subfigure}{0.5\textwidth}
		\includegraphics[width=\textwidth]{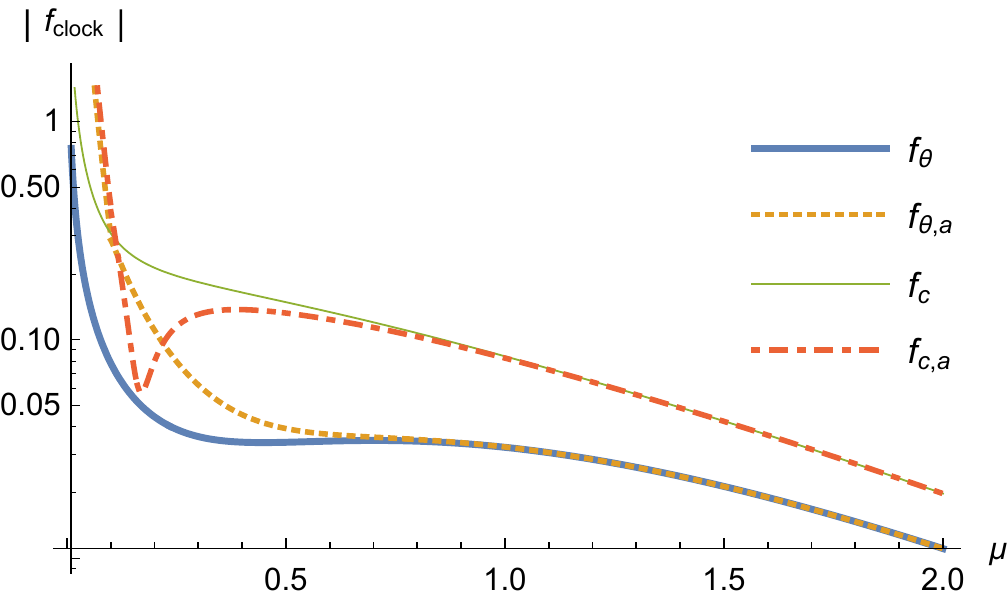}
		\caption{$c=0.1$}
	\end{subfigure}
	\begin{subfigure}{0.5\textwidth}
		\includegraphics[width=\textwidth]{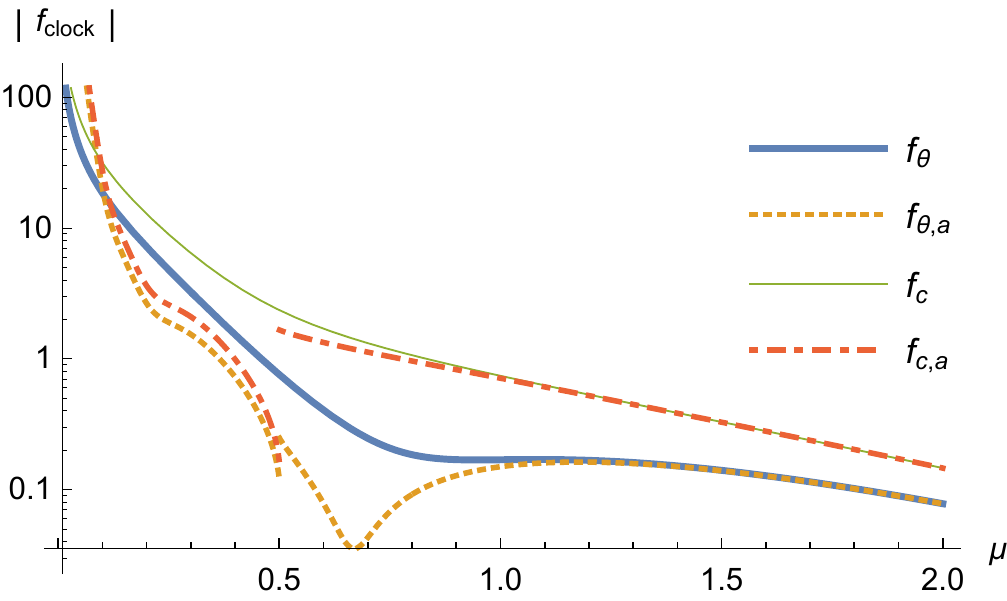}
		\caption{$c=0.5$}
	\end{subfigure}
	\begin{subfigure}{0.5\textwidth}
		\includegraphics[width=\textwidth]{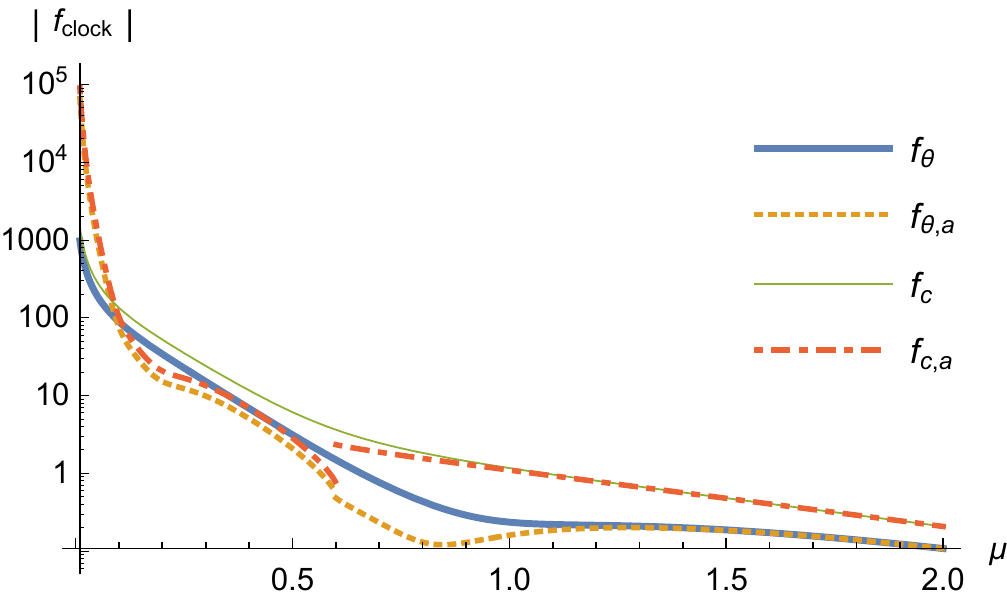}
		\caption{$c=0.6$}
	\end{subfigure}
	\begin{subfigure}{0.5\textwidth}
		\includegraphics[width=\textwidth]{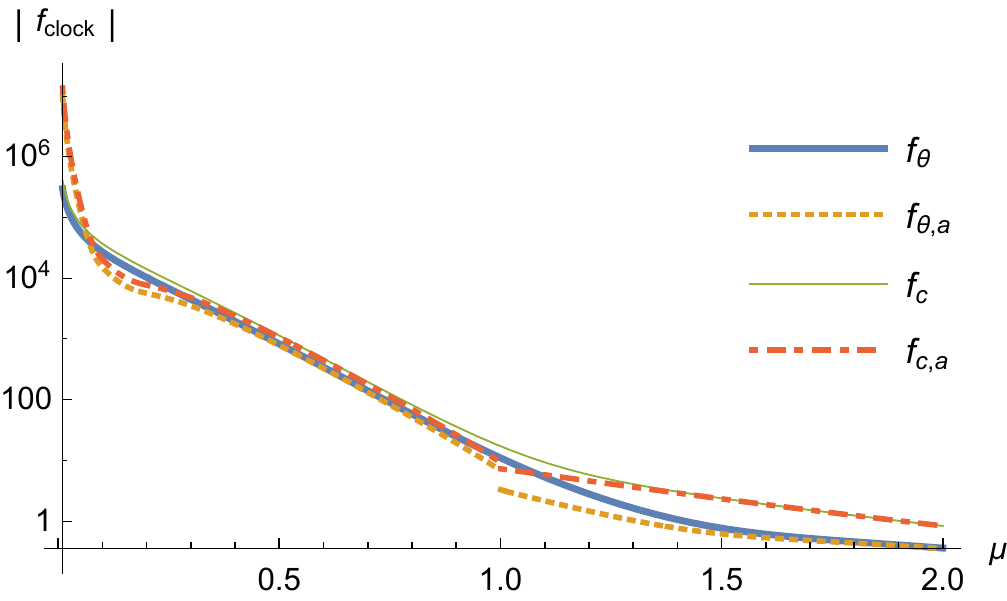}
		\caption{$c=1$}
	\end{subfigure}
\caption{The amplitude of $f_\theta(\mu,c)$ and $f_c(\mu,c)$ defined in \eqref{eqt.fc} from three point function, for $c$ taking different values. The dashed lines show the approximate functions as shown in \eqref{eqt.fthetaa} and \eqref{eqt.fca}, the discontinuities of which take place around $\mu= c$.}
\label{fg.f}
\end{figure*}
Fig.~\ref{fg.f} shows the amplitudes of $f_\theta(\mu,c)$ and $f_c(\mu,c)$, where the solid lines represent the numerical results without taking large $\mu$ limit, and dashed lines represent corresponding approximate results as in \eqref{eqt.fthetaa} and \eqref{eqt.fca}. With these results, we go to the isocurvature bispectrum signals.\black
\section{Bispectrum signal}
\label{sec.bi}
\begin{figure*}[t]\centering
	\begin{subfigure}{0.3\textwidth}
		\includegraphics[width=\textwidth]{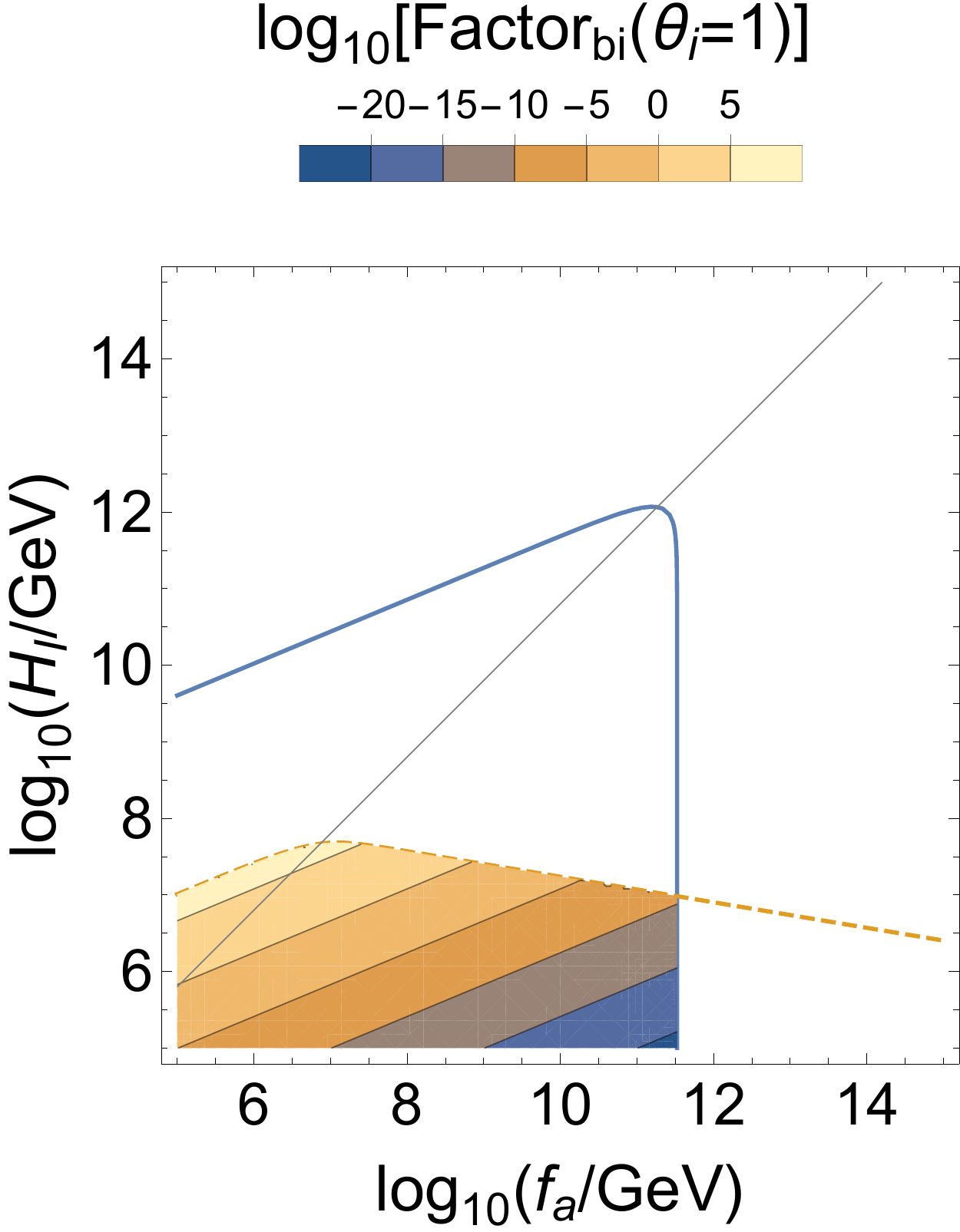}
		\caption{$\theta_i=1$ set in QCD axion model.\phantom{aaa}}
	\end{subfigure}\hspace*{10pt}
	\begin{subfigure}{0.3\textwidth}
		\includegraphics[width=\textwidth]{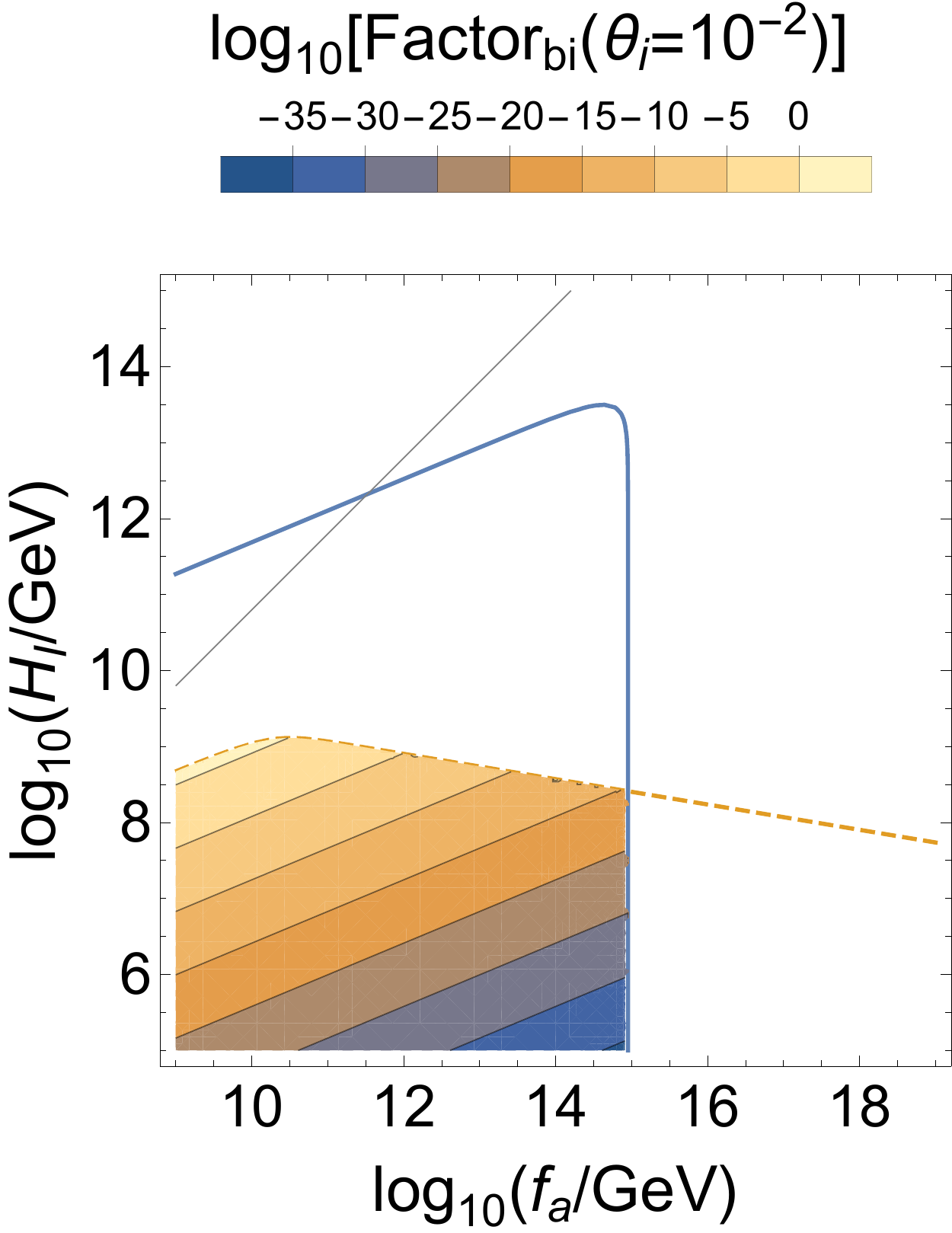}
		\caption{$\theta_i=10^{-2}$ set in QCD axion model.}
	\end{subfigure}\hspace*{10pt}
	\begin{subfigure}{0.31\textwidth}
		\includegraphics[width=\textwidth]{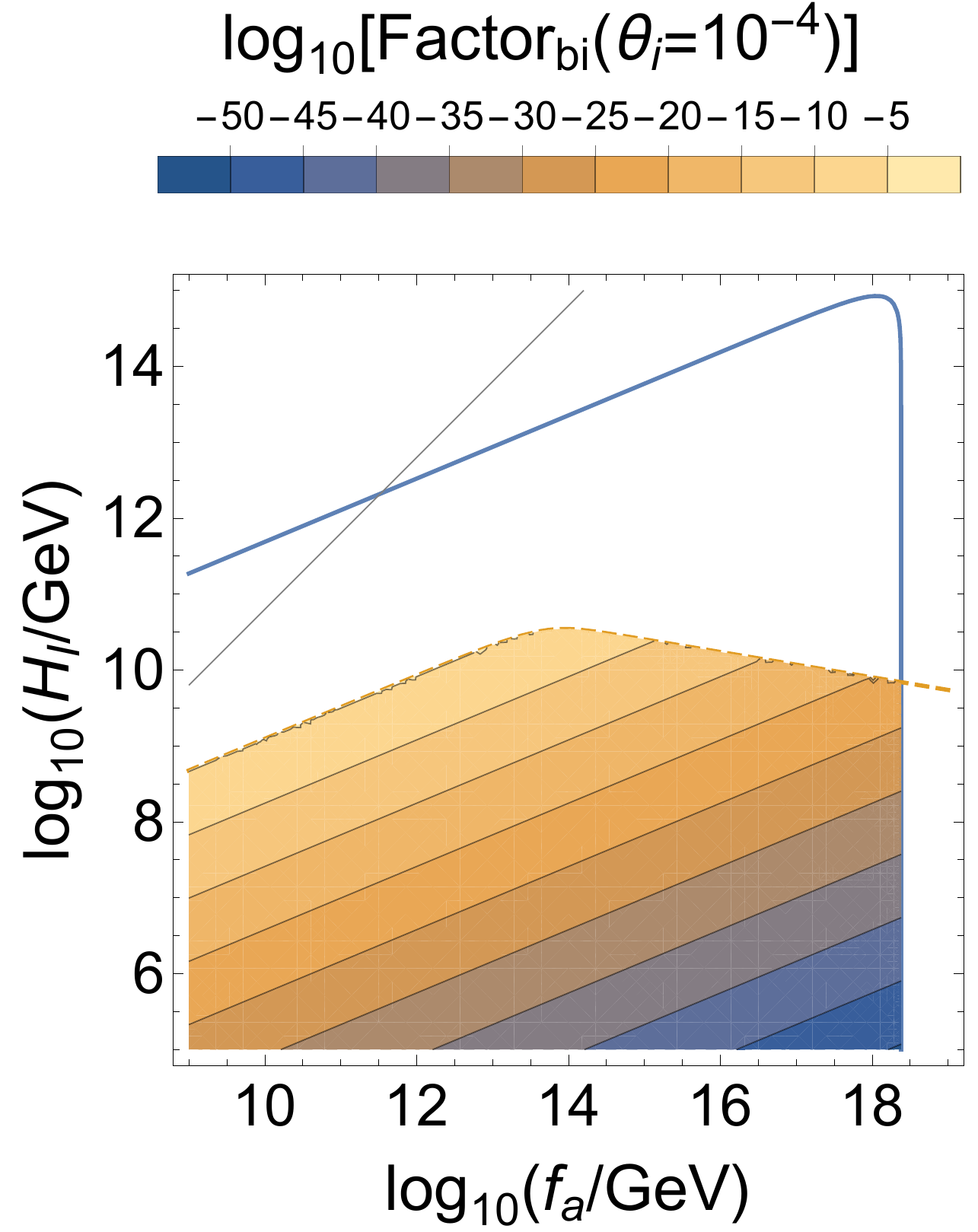}
		\caption{$\theta_i=10^{-4}$ set in QCD axion model.}
	\end{subfigure}
\caption{For QCD axion model in PQ broken scenario, the amplitude of $\text{Factor}_\text{bi}(f_a,H_I)$ defined as \eqref{eqt.factorbi}
\ (to be multiplied to functions displayed in Fig.~\ref{fg.f})\black, is shown in the region constrained by the dark matter abundance and isocurvature spectrum. The solid lines present when axions constitute all dark matter, and dashed lines are from isocurvature constrain, with different initial misalignment angles $\theta_i$, which are the same as shown in Fig.~\ref{fig.QCDiso}. We can see that, the factor in QCD axion case can reach $\mathcal{O}(1)$ when $\theta_i=1$, and maximally of $\mathcal{O}(10^{-5})$ when $\theta_i=10^{-4}$.}
\label{fg.Factorbiqcd}
\end{figure*}

\begin{figure*}[h]\centering
	\begin{subfigure}{0.3\textwidth}
		\includegraphics[width=\textwidth]{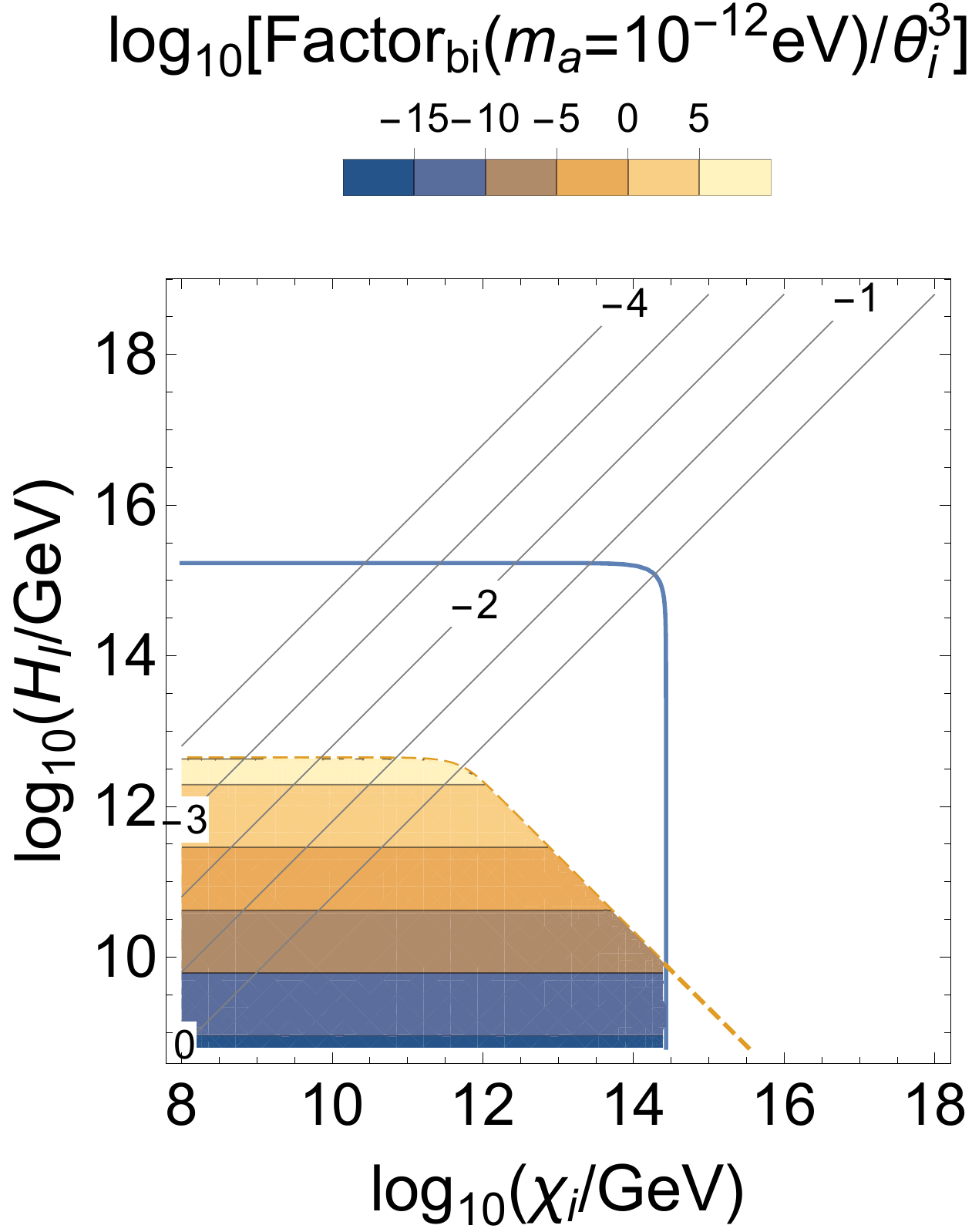}
		\caption{ALP mass $m_a=10^{-12}$eV set.}
	\end{subfigure}\hspace*{10pt}
	\begin{subfigure}{0.3\textwidth}
		\includegraphics[width=\textwidth]{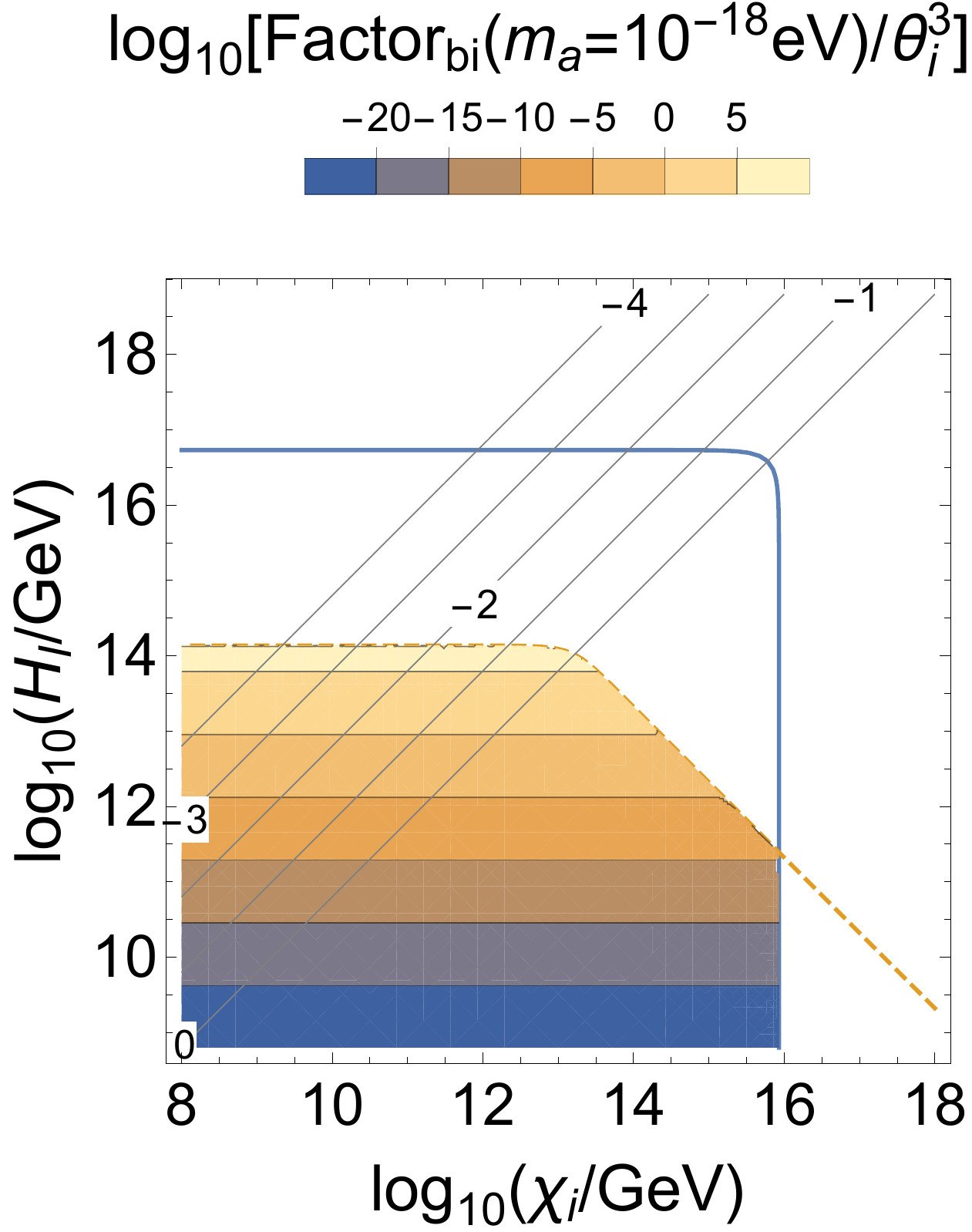}
		\caption{ALP mass $m_a=10^{-18}$eV set.}
	\end{subfigure}\hspace*{10pt}
	\begin{subfigure}{0.3\textwidth}
		\includegraphics[width=\textwidth]{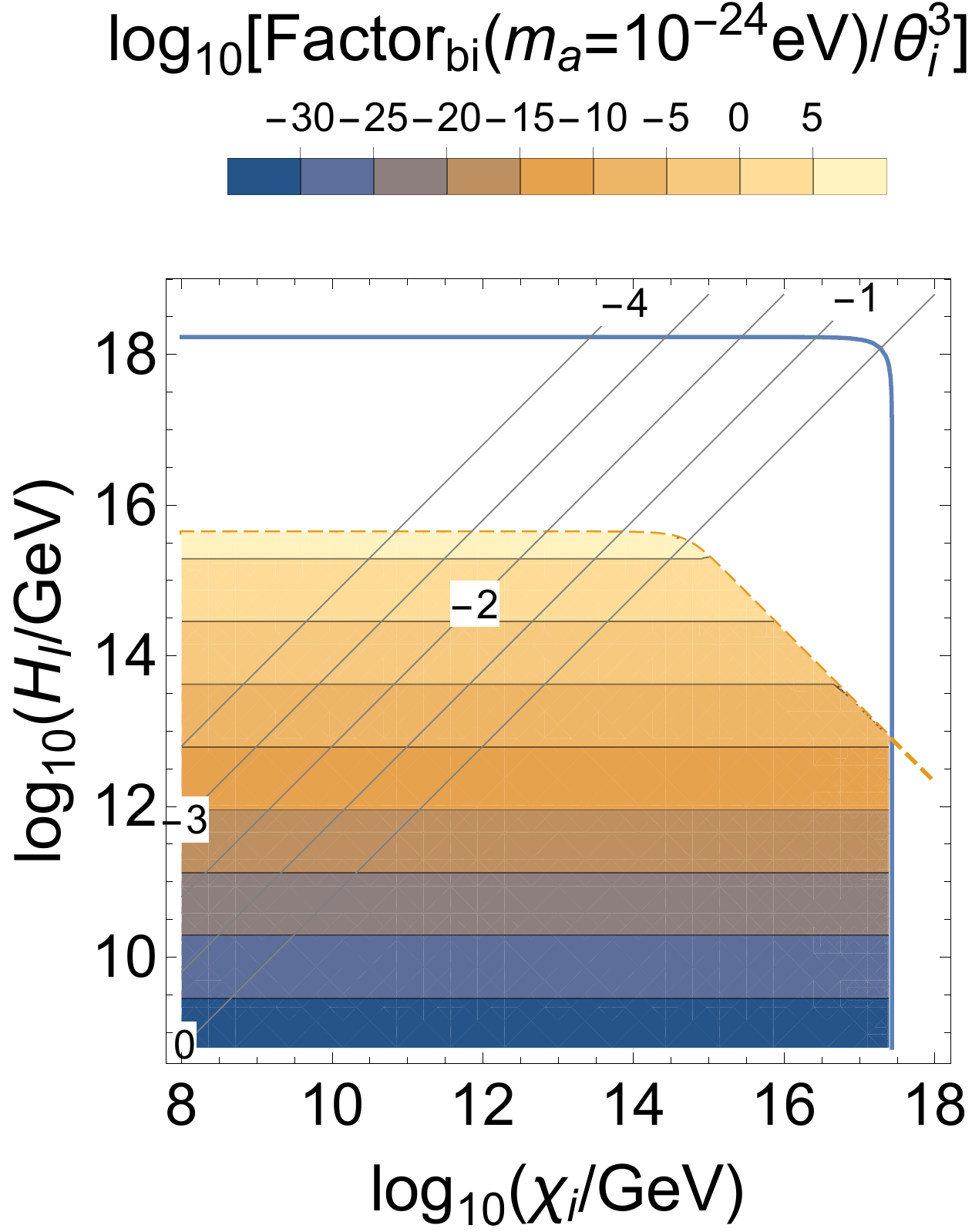}
		\caption{ALP mass $m_a=10^{-24}$eV set.}
	\end{subfigure}
\caption{For PQ broken scenario of ALPs that began oscillation during RD epoch, the amplitude of the factor $[\text{Factor}_\text{bi}(\chi_i\equiv f_a\theta_i,H_I)/\theta_i^3]$ is shown in the region constrained by the dark matter abundance and isocurvature spectrum. The solid lines present when axion constitute all dark matter, and dashed lines are from isocurvature constrain, which are the same as shown in Fig.~\ref{fig.ALPmaiso}. The gray lines show the upper bounds $H_I=2\pi \chi_i/\theta_i$ for PQ broken condition, with different $\log_{10}\theta_i$. We can see that, the factor in ALP model is allowed to reach a large value with reasonable parameters, for cases with different $m_a$.  }
\label{fg.Factorbialp}
\end{figure*}

The multi-field model bispectrums are defined by
\begin{align}
	\langle \zeta^I(\mathbf{k_1})\zeta^J(\mathbf{k_2})\zeta^K(\mathbf{k_3}) \rangle'= B^{IJK }(k_1,k_2,k_3) ~,
\end{align}
where $I,J,K=a\text{ or }i$, for adiabatic mode $\zeta^a=\zeta$ or isocurvature mode $\zeta^i=S$, respectively. 
The axion fluctuations are converted to isocurvature perturbations. According to \eqref{eqt.sa},
\begin{align}
	\langle S_a(\mathbf{k_1})S_a(\mathbf{k_2})S_a(\mathbf{k_3}) \rangle \simeq
\left(\frac{\Omega_a}{\Omega_d}\right)^3 \left(\frac{2\theta_i}{\theta_i^2+\sigma_a^2}\right)^3\langle \delta\theta(\mathbf{k_1})\delta\theta(\mathbf{k_2})\delta\theta(\mathbf{k_3}) \rangle ~,
\end{align}
so we have
\begin{align}
	B_a^{iii}(k_1,k_2,k_3)=\left(\frac{\Omega_a}{\Omega_d}\right)^3\left(\frac{2\theta_i}{\theta_i^2+\sigma_a^2}\right)^3\frac{H_I^6}{f_a^6 k_1^6}\frac{c_0^3 H_I^2}{\pi^2 m_Z^2}\left\{[f_\theta(\mu,c)\cos^2\theta+f_c(\mu,c)]\left(\frac{k_3}{k_1}\right)^{-2i\mu}+\text{c.c.}\right\} ~.
\end{align}
The bispectrum for multi-field model is typically of local shape,
\begin{align}
	B^{iii}(k_1,k_2,k_3)=\frac{2}{5}f^{i,ii}_{\text{NL}}P_I(k_2)P_I(k_3)+\frac{2}{5}f^{i,ii}_{\text{NL}}P_I(k_3)P_I(k_1)+\frac{2}{5}f^{i,ii}_{\text{NL}}P_I(k_1)P_I(k_2) ~,
\end{align}
defines the $f^{i,ii}_{\text{NL}}$, where $P_I(k)=\langle S(\mathbf{k})S(\mathbf{k'})\rangle'$. In the squeezed limit $k_3\ll k_1\sim k_2$, 
\begin{align}
	B_a^{iii}(k_1,k_2,k_3)\simeq \frac{4}{5}f^{i,ii}_{\text{NL},a}P_I(k_3)P_I(k_1)=\frac{(2\pi)^4}{5k_1^3k_3^3}f^{i,ii}_{\text{NL},a}A_I^2 ~,
\end{align}
where the power spectrum amplitude $A_I=A_s\beta_{\text{iso}}$ \footnote{The power spectrum amplitude $A_I(k)=A_I(k/k_0)^{1-n_I}$, where $n_I\simeq 1$. Here we just regard it as scale-independent.}. So
\begin{align}
	f^{i,ii}_{\text{NL},a}=\left(\frac{\Omega_a}{\Omega_d}\right)^3\left(\frac{2\theta_i}{\theta_i^2+\sigma_a^2}\right)^3\frac{5}{(2\pi)^4 A_I^2}\frac{H_I^6}{f_a^6}\frac{c_0^3 H_I^2}{\pi^2 m_Z^2}\left\{[f_\theta(\mu,c)\cos^2\theta+f_c(\mu,c)]\left(\frac{k_3}{k_1}\right)^{3-2i\mu}+\text{c.c.}\right\} ~,
\end{align}
\\
The shape function of bispectrum is defined as
\begin{align}
	\langle  S(\mathbf{k_1})S(\mathbf{k_2})S(\mathbf{k_3}) \rangle'=(2\pi)^4 S(k_1,k_2,k_3)\frac{1}{k_1^2k_2^2k_3^2}A_I^{2} ~,
\end{align}
so we have
\begin{align}
	S(k_1,k_2,k_3)=\left(\frac{\Omega_a}{\Omega_d}\right)^3 \left(\frac{2\theta_i}{\theta_i^2+\sigma_a^2}\right)^3 \frac{1}{(2\pi)^4 A_I^{2}}\frac{H_I^6}{f_a^6}\frac{c_0^3 H_I^2}{\pi^2 m_Z^2}\left\{[f_\theta(\mu,c)\cos^2\theta+f_c(\mu,c)]\left(\frac{k_3}{k_1}\right)^{2-2i\mu}+\text{c.c.}\right\} ~,
\end{align}
\color{black}
where $c_0$ and $({H_I}/{m_Z})$ are typically of $\mathcal{O}(1)$. Therefore, to evaluate the amplitude of axion isocurvature signal, we extract the part
\begin{align}
	\text{Factor}_\text{bi}\equiv\left(\frac{\Omega_a}{\Omega_d}\right)^3\left(\frac{2\theta_i}{\theta_i^2+\sigma_a^2}\right)^3
\left(\frac{H_I}{f_a}\right)^6
\frac{1}{(2\pi)^4 A_I^{2}} ~,\label{eqt.factorbi}
\end{align}
then the final form becomes
\begin{align}
	f^{i,ii}_{\text{NL},a}=\text{Factor}_\text{bi}(\theta_i,m_a;f_a,H_I)\times\left(\frac{5c_0^3H_I^2}{\pi^2m_Z^2}\right)\left\{[f_\theta(\mu,c)\cos^2\theta+f_c(\mu,c)]\left(\frac{k_3}{k_1}\right)^{3-2i\mu}+\text{c.c.}\right\} ~.
\end{align}
Within the constrained parameter space as in Sec.~\ref{sec.constrain}, we have the plots in Fig.~\ref{fg.Factorbiqcd} and Fig.~\ref{fg.Factorbialp} for $\text{Factor}_{\text{bi}}$. We can see that, the factor in QCD axion model is maximally of $\mathcal{O}(1)$ when $\theta_i=1$, and maximally of $\mathcal{O}(10^{-5})$ when $\theta_i=10^{-4}$. While in a ALP model, it is possible to be larger than $\mathcal{O}(10^5)\times\theta_i^3$ for any $\theta_i$. Thus in either case, there exists a parameter space in which the coefficients are not suppressed severely. Meanwhile, the large enough $\text{Factor}_\text{bi}$ with different $\theta_i$ and/or $m_a$ values require different scale constrains on $(f_a,H_I)$. For example, taking a non fine-tuning initial misalignment angle $\theta_i\simeq 1$ in QCD axion model, the inflation is constrained to be of low energy $H_I<10^{8}$ GeV, and a $\text{Factor}_\text{bi}>\mathcal{O}(10^{-5})$ requires axion decay constant $f_a\lesssim 10^{9}$ GeV. While $\theta_i\simeq 1$ in ALP model allows higher $H_I\simeq 10^{11\sim 15}$ GeV and correspondingly higher $f_a\simeq 10^{10\sim 16}$ GeV under different ALP masses $m_a=10^{-24}\sim 10^{-12}$ eV, for a factor $\text{Factor}_\text{bi}\simeq 10^{-5} \sim 10^5$. Smaller $\theta_i$ lead to much smaller bispectrum signals permitted \footnote{Note that we are assuming ALPs here began oscillation during RD epoch, which need to be checked for specific models, especially for heavier ALPs.}.
\color{black}
\begin{remark}[]\small
	We can analyze the plots by taking limits $\sigma_a\ll\theta_i$ and $\theta_i\ll\sigma_a$. 
	\begin{itemize}
		\item QCD axion model (PQ broken scenario condition equivalent to $\sigma_a<1$). When $\sigma_a\gg\theta_i$, we have
			\begin{align}\left\{
			\begin{array}{rl}
				\log\Omega_a\sim & 2\log H_I-\frac{5}{6}\log f_a\\ 
				\log\alpha\sim & 2(2\log H_I-\frac{5}{6}\log f_a)\\
				\log\text{Factor}_\text{bi}\sim & 3(2\log H_I-\frac{5}{6}\log f_a)+3\log\theta_i
			\end{array} \right. ~,
			\end{align}
		while when $\theta_i\gg\sigma_a$,
			\begin{align}\left\{
			\begin{array}{rr}
				\log\Omega_a\sim & \frac{7}{6}\log f_a+2\log\theta_i\\ 
				\log\alpha\sim & 2\log H_I+\frac{1}{3}\log f_a+2\log\theta_i\\
				\log\text{Factor}_\text{bi}\sim & 6\log H_I-\frac{5}{2}\log f_a+3\log\theta_i
			\end{array} \right. ~.
			\end{align}
		We can see that the plots obey these well in almost whole region except the turning point $\sigma_a=\theta_i$. So $\text{Factor}_\text{bi}$ reach maximum on the boundary by $\alpha=0.038$ and $\sigma_a\lesssim\theta_i$.
		\item ALP model (PQ broken scenario condition equivalent to $\sigma_a<1$). When $\sigma_a\gg\theta_i$, we have
			\begin{align}\left\{
			\begin{array}{rl}
				\log\Omega_a\sim & 2\log H_I+\frac{1}{2}\log m_a\\ 
				\log\alpha\sim & 4\log H_I+\log m_a\\
				\log(\text{Factor}_\text{bi}/\theta_i^3)\sim & 6\log H_I+\frac{3}{2}\log m_a
			\end{array} \right. ~,
			\end{align}
		while when $\theta_i\gg\sigma_a$,
			\begin{align}\left\{
			\begin{array}{rl}
				\log\Omega_a\sim & \phantom{2\log H_I+}2\log\chi_i+\frac{1}{2}\log m_a\\ 
				\log\alpha\sim & 2\log H_I+2\log\chi_i+\log m_a\\
				\log(\text{Factor}_\text{bi}/\theta_i^3)\sim & 6\log H_I\phantom{+2\log\chi_i}+\frac{3}{2}\log m_a
			\end{array} \right. ~.\label{eqt.alpbiestim}
			\end{align}
		We can see that the plots obey these well in almost whole region except the turning point $\sigma_a=\theta_i$. So $\text{Factor}_\text{bi}$ reach maximum on the boundary by $\alpha=0.038$ and $\sigma_a\lesssim\theta_i$.
	\end{itemize}
\end{remark}
\color{black}\normalsize
\indent
Combined with Fig.~\ref{fg.f} for $f_\theta(\mu,c)$ and $f_c(\mu,c)$, we can see that, under the constrains, QCD axion interacting to massive boson during inflation can contribute large non-Gaussian signal at lower scale inflation without misalignment fine-tuning. While ALP model can give large non-Gaussian signal at higher scale inflation without misalignment fine-tuning. 
\\ \indent
In QCD axion model, for example, 
$
	\text{take }
\theta_i=1,\ f_a=10^9~\text{GeV},\ H_I=10^{7.4}~\text{GeV},
	\text{we have }
$
\begin{align}\nonumber
	 \text{Factor}_\text{bi}=0.27,\ \text{and}\ \alpha=0.037,\ \Omega_a/\Omega_d=1.1\times 10^{-3},\ \sigma_a=4.0\times 10^{-3},
\end{align}
	and take
$
\mu=0.1,\ c=0.5,\ c_0=1,
$
	we have
\begin{align}\nonumber
	m_Z/H_I=0.51,\ |f_\theta(\mu,c)|=18.5,\ &|f_c(\mu,c)|=31.4 ~,\\
\nonumber
	\frac{5c_0^3H_I^2}{\pi^2m_Z^2}=1.95,\ f_\theta(\mu,c)=10.5+15.2\ii,\ &f_c(\mu,c)=15.0+27.6\ii ~,
\end{align}
final result
\begin{align}\nonumber
	f_{\text{NL},a}^{i,ii}=1.05\left\{(15.0+10.5\cos^2\theta)\cos\left[0.2\log\left({k_3}/{k_1}\right)\right]\right. \left.+(27.6+15.2\cos^2\theta)\sin\left[0.2\log\left({k_3}/{k_1}\right)\right]\right\}\left({k_3}/{k_1}\right)^3 ~.
\end{align}
For ALP model,
$
	\text{take }
m_a=10^{-18}~\text{eV},\ \theta_i=1,\ \chi_a=10^{13.8}~\text{GeV},\ H_I=10^{13.5}~\text{GeV},
	\text{we have }
$
\begin{align}\nonumber
	 \text{Factor}_\text{bi}=1.9\times 10^3,\ \text{and}\ \alpha=0.035,\ \Omega_a/\Omega_d=5.5\times 10^{-5},\ \sigma_a=0.08,
\end{align}
	and take
$
\mu=0.5,\ c=0.5,\ c_0=1,
$
	we have
\begin{align}\nonumber
	m_Z/H_I=0.71,\ |f_\theta(\mu,c)|=0.75,\ &|f_c(\mu,c)|=2.34 ~,\\
\nonumber
	\frac{5c_0^3H_I^2}{\pi^2m_Z^2}=1.01,\ f_\theta(\mu,c)=0.67-0.34\ii,\ &f_c(\mu,c)=2.26+0.64\ii ~,
\end{align}
final result
\begin{align}\nonumber
	f_{\text{NL},a}^{i,ii}=3.82\times 10^{3}\left\{(2.26+0.67\cos^2\theta)\cos\left[\log\left({k_3}/{k_1}\right)\right]\right. \left.+(0.64-0.34\cos^2\theta)\sin\left[\log\left({k_3}/{k_1}\right)\right]\right\}\left({k_3}/{k_1}\right)^3 ~.
\end{align}
\paragraph {Discussions}
	We have discussed our isocurvature bispectrum signal, and calculated some examples with $c\sim\mathcal{O}(0.1)$. However, considering the axion models above, the parameter $c=({\dot\theta_0}/{H_I})c_0$ is dependent on the rolling speed during inflation. According to the equation of motion \eqref{eqt.chieom}, $\dot\theta_0=-({m_a^2}/{H_I})\theta_0$ can be severely suppressed for axions constituting DM with $m_a\ll H_I$. We can see that axion DM realigned after EW breaking have $m_a<1~\text{eV}$, and general ALP and QCD axion models give $f_a>10^9~\text{GeV}$. The not-small bispectrum signals do not allow very low scale $H_I$, according to \eqref{eqt.alpbiestim} and Fig.~\ref{fg.Factorbialp} for $\text{Factor}_{bi}$. Therefore, $c$ is extremely small in this case, thus making the whole bispectrum signal vanish through $f_\theta$ and $f_c$~\footnote{Investigate the case with extremely small $c\rightarrow 0$, we can see that only $I_{AA}$ in \eqref{eqt.integral} survives, which was a subleading term for squeezed limit $k_3/k_1$ order when all terms not vanished. But this term still gives a zero result after loop integral $\int^{2\pi}_0 d\varphi_1$.}. Axiverse model with dynamical modulus $\sigma_i$ may solve this problem, giving the $m_a\propto e^{-\#\sigma_i}$ and decrease during $\sigma_i$ field evolution\cite{Marsh:2011gr}. Thus this kind of ALP models may hopeful provide light enough $m_a$ for axion DM, while do not conflict with a large enough $c$ during inflation, as taken in above examples.
\\
\indent

\section{Conclusion}
Axions can serve as cosmological colliders for particles interacting with them. By studying the information left on axion fluctuations, we may find signature of these intermediate particles, and get their properties. For light axions that constitute cold DM, the fluctuations are reflected on primordial perturbations as isocurvature modes. In this paper, we considered such axion isocurvature colliders. Though quite dependent on late time evolution, this mechanism is still plausible to give significant characteristic clock signals, in some axion models. We discussed parameter space required by such models, under the constrains by CDM relic abundance and isocurvature power spectrum. 
And showed that the contribution to isocurvature non-Gaussianity $f_{\text{NL}}^{i,ii}$ can possibly reach a large value $\mathcal{O}(10^2)$. 
\\ \indent
Two types of models are usually considered, QCD axions and ALPs, the non-perturbative effects of which are switched on at different energy scales, and the axion masses dependent differently on decay constant. ALP models are mostly established from high-dimensional theories, which are of rich phenomenology, thus quite flexible. We investigated both types under the PQ broken scenario, taken a boson as the intermediate particle, and showed that, the leading bispectra from axion coupling can be written and evaluated in two parts. One from an inflationary process with the same form, and the other is dependent on the different evolutions after inflation in different models. The information of intermediate particle and angular dependence are in the former part, while the latter one multiplies such signatures. We showed that, both parts can reach significant values under reasonable parameters, both for QCD axions and ALPs. While for QCD axions, the constrains are more tightened, going close to the lower bound of decay constant $f_a$ from BHSR if we do not like a too tiny initial misalignment angle $\theta_i$ by PQ broken. We have not considered specific ALP models, and just showed results at different fixed ULA masses $m_a$. With these, we can already see that, ALPs can be much more free to realize a large bispectrum signal. Another problem has been discussed above, that the light axion $m_a<1~\text{eV}\ll H_I$ may conflict with the requirement, that $c$ should not be small or the signature part will vanish. Mechanisms such as dynamical moduli in ALP models can solve the confliction through a dynamical $m_a$. We then took numerical examples to show the results. 
\\ \indent
Axion model realizations are quite rich. We did not study in very details about QCD axions or ALPs. As a general mechanism, axion field in current patch got an initial misalignment angle $\theta_i$ from the broken PQ symmetry during inflation, then keep slow-rolling and behaved like dark energy, before Hubble parameter goes to be comparable to axion mass. After oscillation began at $m_a\sim H$, we regard axions as CDM till today. The axions of our concerned masses mostly began oscillation during RD epoch. We also only consider the impact from axions, neglecting other possible matter produced by particular axion models. Besides, only the axion production through misalignment channel enters our final results. Furthermore, details of dynamical ALPs or other solutions for the slow-roll conflict are interesting to be considered. Other types of intermediate particles are also wealthy to be investigated. These can all be left for further works. Our work is to show the possibility of realization.
\section*{Acknowledgement}
We thank Yi Wang, Zhong-Zhi Xianyu and Yue Zhao for the nice discussions. 
\appendix
\section{The whole result of indices contraction}\label{sec.appendixA}
\begin{align}
\nonumber I^{\ ba}_{AA}
	=\{[\partial_\tau A(p)\partial_{\tau'} A(q) \partial_\tau
+ i A(p)\partial_{\tau'} A(q) (\mathbf{p}\cdot i\mathbf{k_1})]G(k_1)(\mathbf{p\times q}\cdot i\mathbf{k_2})G(k_2)+(\mathbf{p}\leftrightarrow\mathbf{q})\}&\\
+(\mathbf{k_1}\leftrightarrow\mathbf{k_2})
\end{align}
\begin{align}
\nonumber
	I^{\ ba}_{AB}=\{[\partial_{\tau'}\partial_{\tau} B(p) A(q) (\mathbf{p\cdot q}) \partial_{\tau} - 
	     i \partial_{\tau'}\partial_{\tau} B(p)\partial_{\tau} A(q)(\mathbf{p}\cdot i\mathbf{k_1})]G(k_1)(\mathbf{p\times q\cdot}i\mathbf{k_2})G(k_2)&\\ \nonumber
+(\mathbf{p}\leftrightarrow\mathbf{q})\}&\\
+(\mathbf{k_1}\leftrightarrow\mathbf{k_2})
\end{align}
\begin{align}
	I^{\ ba}_{BB}=0
\end{align}
\begin{align}
\nonumber
	 I^{\ ba}_{BC}=\{\underline{+i \partial_{\tau} C(p) \partial_{\tau'} \partial_{\tau}B(q)}&\underline{[(\mathbf{p\times q\cdot}i\mathbf{k_1})(\mathbf{p\times q\cdot}i\mathbf{k_2}) - (\mathbf{p\cdot q}) (\mathbf{p}\cdot i\mathbf{k_1}) (\mathbf{q}\cdot i\mathbf{k_2})} \\ \nonumber
    &\underline{+ q^2(\mathbf{p}\cdot i\mathbf{k_1})(\mathbf{p}\cdot i\mathbf{k_2})]}
\underline{G(k_1)G(k_2)}\\ \nonumber 
	+C(p)\partial_{\tau'}\partial_{\tau}B(q) p^2 &[- q^2(\mathbf{p}\cdot i\mathbf{k_1})+  (\mathbf{p\cdot q})(\mathbf{q}\cdot i\mathbf{k_1})]G(k_1)\partial_{\tau}G(k_2)
+(\mathbf{p}\leftrightarrow\mathbf{q})\}\\
&\hspace*{210pt}+(\mathbf{k_1}\leftrightarrow\mathbf{k_2})
\end{align}
\begin{align}
\nonumber
	I^{\ ba}_{CC}=( \underline{ \{ } &\underline{ [i \partial_{\tau'}\partial_{\tau}C(p) \partial_{\tau}C(q) } + i C(p)\partial_{\tau'} C(q) p^2\underline{ ](\mathbf{q}\cdot i\mathbf{k_1}) }  \\ \nonumber
	+ &[-\partial_{\tau'}\partial_{\tau}C(p) C(q) q^2 - \partial_{\tau'} C(p)\partial_{\tau} C(q)(\mathbf{p\cdot q})]\partial_{\tau}
\underline{\}
G(k_1)(\mathbf{p\times q\cdot}i\mathbf{k_2})G(k_2)} \\
	&\hspace*{210pt}+(\mathbf{p}\leftrightarrow\mathbf{q}) )
+(\mathbf{k_1}\leftrightarrow\mathbf{k_2})
\end{align}
\begin{align}
\nonumber
	I^{\ ba}_{AC}=
	\{\underline{+i \partial_{\tau'}\partial_{\tau}C(p)\partial_{\tau}A(q)}&\underline{[-(\mathbf{p}\cdot i\mathbf{k_1}) (\mathbf{q}\cdot i\mathbf{k_2})  - (\mathbf{p\cdot q})(i \mathbf{k_1}\cdot i\mathbf{k_2})]G(k_1)G(k_2)}
\\ \nonumber
	\underline{+i \partial_{\tau}C(p) \partial_{\tau'}\partial_{\tau}A(q)}&\underline{[ p^2 (i \mathbf{k_1}\cdot i\mathbf{k_2}) + (\mathbf{p}\cdot i\mathbf{k_1})(\mathbf{p}\cdot i\mathbf{k_2})]G(k_1)G(k_2)}
\\ \nonumber
	+ i \partial_{\tau'}C(p) A(q)& [-(\mathbf{p\times q\cdot}i\mathbf{k_1})(\mathbf{p\times q\cdot}i\mathbf{k_2}) + (\mathbf{p\cdot q}) (\mathbf{p}\cdot i\mathbf{k_1}) (\mathbf{q}\cdot i\mathbf{k_2})\\ \nonumber
&-  (\mathbf{p\cdot q})^2 (i \mathbf{k_1}\cdot i\mathbf{k_2})]G(k_1)G(k_2)
\\ \nonumber
	+ i C(p) \partial_{\tau'}A(q) p^2 &[(\mathbf{p\cdot q})(i \mathbf{k_1}\cdot i\mathbf{k_2}) - (\mathbf{p}\cdot i\mathbf{k_1})(\mathbf{q}\cdot i\mathbf{k_2})]G(k_1) G(k_2)
\\ \nonumber
	+\partial_{\tau'}\partial_{\tau} C(p) A(q)& [(\mathbf{p\cdot q}) (\mathbf{q}\cdot i\mathbf{k_1}) + q^2 (\mathbf{p}\cdot i\mathbf{k_1})]G(k_1)\partial_{\tau}G(k_2)
\\ \nonumber
+ C(p)\partial_{\tau'}\partial_{\tau} A(q)&[-2 (\mathbf{p}\cdot i\mathbf{k_1}) p^2]G(k_1)\partial_{\tau}G(k_2) 
\\ \nonumber
	+\partial_{\tau} C(p)\partial_{\tau'} A(q)& [-(\mathbf{p}\cdot i\mathbf{k_1}) (\mathbf{p\cdot q}) - p^2(\mathbf{q}\cdot i\mathbf{k_1})]G(k_1)\partial_{\tau}G(k_2)
\\ \nonumber
+ \partial_{\tau'} C(p) \partial_{\tau}A(q) &[+(\mathbf{p}\cdot i\mathbf{k_1}) (\mathbf{p\cdot q}) + p^2(\mathbf{q}\cdot i\mathbf{k_1})]G(k_1)\partial_{\tau}G(k_2)
\\ \nonumber
	+ i \partial_{\tau'}C(p) A(q)& [(\mathbf{p\cdot q})^2 + p^2 q^2] \partial_{\tau}G(k_1)\partial_{\tau}G(k_2)
\\ \nonumber
	+i C(p) \partial_{\tau'}A(q) p^2 &[ - 2 (\mathbf{p\cdot q})] \partial_{\tau}G(k_1)\partial_{\tau}G(k_2)
\\&\hspace*{70pt}+(\mathbf{p}\leftrightarrow\mathbf{q})\}
+(\mathbf{k_1}\leftrightarrow\mathbf{k_2}) ~,
\end{align}
here we omitted some parameters to make the equations look less complicated, e.g. use $A(p)$ to denote $A^{ba}(p,\tau',\tau)$ and $G(k)$ to denote $G^{a}(k,\tau)$. \\ \indent
The underlined are leading-order terms in the squeezed limit $k_1\simeq k_2 \gg k_3$, with $p=q=k_3$. This can be obtained by
\begin{align}
	A(p)\sim p^2B(p)&\sim p C(p)\sim (\tau'\tau)^{1/2} ~,\\
	\partial\tau\sim \tau^{-1}&,\ \partial\tau'\sim \tau'^{-1} ~,
\end{align}
and due to the time integral
\begin{align}
	\tau\sim k_1^{-1},\ \tau' \sim k_3^{-1} ~,
\end{align}
where the sim symbol denotes equivalences for $(k_3/k_1)$-order. With these, we can see that, terms with the form like $\partial_{\tau'}\partial_{\tau}(A,B,\text{or }C)
\times\partial_\tau(A,B,\text{or }C)$ have the maximal order in the squeezed limit.
\section{Details of integral}\label{sec.appendixB}
\subsection{Late time expansion}
At late time limit $(-p\tau)\rightarrow 0$, the mode function
\begin{align}\nonumber
	v^{\scriptscriptstyle{(+)}}_p(\tau)\simeq 
e^{-i\pi/4}(-\tau)^{1/2}&\left[f(\mu,c)(-p\tau)^{i\mu}+f(-\mu,c)(-p\tau)^{-i\mu}
\right] ~,
\end{align}
and
\begin{align}
	v^{\scriptscriptstyle{(-)}}_p(\tau)=v^{\scriptscriptstyle{(+)}}_p(\tau)|_{c\rightarrow -c} ~,\\
	v^{\scriptscriptstyle{(\parallel)}}_p(\tau)=v^{\scriptscriptstyle{(+)}}_p(\tau)|_{c\rightarrow 0} ~,
\end{align}
for real $\mu$ and $c$, where $f(\mu,c)=\frac{\Gamma(-2i\mu)}{\Gamma(1/2-i\mu-ic)}e^{\pi(\mu-c)/2}2^{i(\mu-c)}$. Then
\begin{align}\nonumber
	u^{\scriptscriptstyle(+)}_p(\tau ',\tau) =v^{\scriptscriptstyle(+)}_p(\tau')(v^{\scriptscriptstyle(+)}_p(\tau))&^* \\
	\simeq (-\tau')^{1/2}(-\tau)^{1/2}&\left[g_1(\mu,c)(-p\tau')^{i\mu}(-p\tau)^{i\mu}+\right. 
\left.g_2(\mu,c)(-p\tau')^{-i\mu}(-p\tau)^{i\mu}+(\mu\rightarrow -\mu)
\right] ~,
\end{align}
where $g_1(\mu,c)=f(\mu,c)f(-\mu,c)^*$ and $g_2(\mu,c)=f(-\mu,c)f(-\mu,c)^*$,
\begin{align}\nonumber
	A^{\scriptscriptstyle{-+}}(p,\tau ',\tau)\simeq (\tau'\tau)^{1/2}&\left[A_1(\mu,c)(-p\tau')^{i\mu}(-p\tau)^{i\mu}+A_2(\mu,c)(-p\tau')^{-i\mu}(-p\tau)^{i\mu}+(\mu\rightarrow -\mu)
\right] ~,\\ \nonumber
	B^{\scriptscriptstyle{-+}}(p,\tau ',\tau)\simeq \frac{1}{p^2}(\tau'\tau)^{1/2}&\left[B_1(\mu,c)(-p\tau')^{i\mu}(-p\tau)^{i\mu}+B_2(\mu,c)(-p\tau')^{-i\mu}(-p\tau)^{i\mu}+(\mu\rightarrow -\mu)
\right] ~,\\
	C^{\scriptscriptstyle{-+}}(p,\tau ',\tau)\simeq \frac{1}{p}(\tau'\tau)^{1/2}&\left[C_1(\mu,c)(-p\tau')^{i\mu}(-p\tau)^{i\mu}+C_2(\mu,c)(-p\tau')^{-i\mu}(-p\tau)^{i\mu}+(\mu\rightarrow -\mu)
\right] ~,
\end{align}
where 
\begin{align} \nonumber
	A_{1,2}(\mu,c)&=\frac{1}{2}[g_{1,2}(\mu,c)+g_{1,2}(\mu,c)|_{c\rightarrow -c}] ~, 
\\ \nonumber
	B_{1,2}(\mu,c)&=g_{1,2}(\mu,0)-\frac{1}{2}[g_{1,2}(\mu,c)+g_{1,2}(\mu,c)|_{c\rightarrow -c}] ~,
\\ 
	C_{1,2}(\mu,c)&=\frac{-i}{2}[g_{1,2}(\mu,c)-g_{1,2}(\mu,c)|_{c\rightarrow -c}] ~.
\end{align}
Then time integral all have the form as
\begin{align}
	\int^0_{-\infty}\text{d}\tau\int^0_{-\infty}\text{d}\tau'\tau'^r\tau^s e^{ibk_3\tau'}e^{iak_{12}\tau}= \Gamma(1+r)\Gamma(1+s)(-iak_{12})^{-1-s}(-ibk_3)^{-1-r}~,
\end{align}
where $\text{Re}(r+1)>0$ $\text{Re}(s+1)>0$.
With these we can work out the time integral as \eqref{eqt.IBCtime}, \\
We can write
\begin{align}\nonumber
	A_2(\mu,c)& =\frac{2 e^{-2 \pi \mu}+e^{-2 \pi  c}+e^{2 \pi c}}{4\mu (e^{2 \pi  \mu }-e^{-2 \pi \mu}) } ~,\\ \nonumber
	B_2(\mu,c)& =\frac{2-e^{2 \pi  c}-e^{- 2 \pi c}}{4\mu \left(e^{2 \pi  \mu }-e^{-2 \pi  \mu }\right) } ~,\\ \nonumber
	C_2(\mu,c)& =\frac{i \left(e^{2 \pi  c}-e^{-2 \pi  c}\right)}{4\mu \left(e^{2 \pi  \mu }-e^{-2 \pi  \mu }\right)} ~,\\
\end{align}
and
\begin{align}\nonumber
	A_1(\mu,c)=& \frac{2^{2 i \mu } \Gamma (-2 i \mu )^2  \cosh (\pi  c)}{\Gamma \left(\frac{1}{2}-i c-i \mu \right) \Gamma \left(\frac{1}{2}+i c-i \mu \right)}\\ \nonumber
	B_1(\mu,c)=& 2^{2 i \mu } \Gamma (-2 i \mu )^2 \left[\frac{1}{\Gamma \left(\frac{1}{2}-i \mu \right)^2}-\frac{\cosh (\pi  c)}{\Gamma \left(\frac{1}{2}-i c-i \mu \right) \Gamma \left(\frac{1}{2}+i c-i \mu \right)}\right] \\ \nonumber
	C_1(\mu,c)=& i\frac{   2^{2 i \mu } \Gamma (-2 i \mu )^2 \sinh(\pi c)}{\Gamma \left(\frac{1}{2}-i c-i \mu \right) \Gamma \left(\frac{1}{2}+i c-i \mu \right)}\\ 
\end{align}
\subsubsection{Large $\mu$ limit}
\label{sec.appendixBmu}
Gamma functions can be expanded into exponential form \footnote{Actually $|y|$ do not have to go to infinity, the approximation has the right order for $|y|\geq\mathcal{O}(0.1)$.}
\begin{align}
	\Gamma(iy)\xrightarrow{|y|\rightarrow\infty} \sqrt{2 \pi } |y|^{-1/2} e^{-\pi |y|/2 } e^{i [y  ( \log | y | -1) -\pi\ \text{sgn}(y)/4 )]}
\end{align}
where $y$ is real. So we have
\begin{align}\nonumber
	A_1(\mu,c)
\xrightarrow[|\mu-c|,|\mu+c|\rightarrow\infty]{|\mu|\rightarrow\infty}& 4^{-1}|\mu|^{-1}\left(e^{\pi  c}+e^{-\pi  c}\right) e^{ \pi  (|\mu-c| +|\mu+c| -4 |\mu| )/2}\times\\
& \times e^{i [(\mu -c) \log |\mu-c|+(\mu+c) \log|\mu+c| -4 \mu  \log |\mu| )]}e^{i[2 \mu (1-\log 2)+\pi  \text{sgn}(\mu )/2]} ~,
\\ \nonumber
	B_1(\mu,c)
	\xrightarrow[|\mu-c|,|\mu+c|\rightarrow\infty]{|\mu|\rightarrow\infty}& 4^{-1}|\mu|^{-1} \{2 e^{-\pi  \left| \mu \right| } -(e^{\pi c}+e^{-\pi  c}) e^{ \pi (|\mu-c| + |\mu+c| -4 |\mu|)/2}\times\\ 
& \times e^{i [(\mu -c) \log |\mu-c|+(\mu+c) \log|\mu+c| -2 \mu  \log |\mu| )]} \}e^{ i[2 \mu  (1-\log |2\mu|)+\pi  \text{sgn}(\mu )/2]} ~,
\\ \nonumber
	C_1(\mu,c)
\xrightarrow[|\mu-c|,|\mu+c|\rightarrow\infty]{|\mu|\rightarrow\infty}& i 4^{-1} |\mu|^{-1} \left(e^{\pi  c}-e^{-\pi c}\right) e^{ \pi  (|\mu-c| +|\mu+c| -4 |\mu| )/2}\times\\
& \times e^{i [(\mu -c) \log |\mu-c|+(\mu+c) \log|\mu+c| -4 \mu  \log |\mu| )]}e^{i[2 \mu (1-\log 2)+\pi  \text{sgn}(\mu )/2]} ~.
\end{align}
With these we can get an approximate result in the form of exponential functions.
\subsection{Contribution by each part}
The result of leading order will have a form like
\begin{align}
	\sum_{s=\scriptscriptstyle{BC,CA,AC}}F^s(\cos\theta)\times \frac{H^6}{f^6 k_1^6 k_3^3} \left[ f^s_1(\mu ,c)(k_1/k_3)^{2 i \mu }+f^s_2(\mu ,c)+ f^s_3(\mu ,c)(k_1/k_3)^{-2 i \mu }\right]
\end{align}
where $f^s_{1,2,3}(\mu,c)$ are from time integral, and $F^s(\cos\theta)$ are from the integration of angle $\varphi_1$.
\paragraph{Contribution by $I^{\ ba}_{BC,(0)}$}
The contribution to \eqref{eqt.diagram}
\begin{align}\nonumber
	&\frac{4c^3_0H^2}{m^2_Z}\int\frac{\mathbf{d}^3\mathbf{p}}{(2\pi)^3}k_1^{-2}k_3^{-4}i\times\\ \nonumber
	&\times\{[(\mathbf{p\times q\cdot}i\mathbf{k_1})(\mathbf{p\times q\cdot}i\mathbf{k_2})-(\mathbf{p\cdot q}) (\mathbf{p}\cdot i\mathbf{k_1}) (\mathbf{q}\cdot i\mathbf{k_2})+ q^2(\mathbf{p}\cdot i\mathbf{k_1})(\mathbf{p}\cdot i\mathbf{k_2})+(\mathbf{p}\leftrightarrow\mathbf{q})]\\ \nonumber
	&\hspace*{345pt}+(\mathbf{k_1}\leftrightarrow\mathbf{k_2})\} 
\\ \nonumber
\simeq &
	\frac{4c^3_0H^2}{m^2_Z}\frac{k_3^3}{(2\pi)^3}\int^{2\pi}_0 d\varphi_1\ i\times\\
&\times \bigg[(3\sin^2\theta\sin^2\varphi_1)+\frac{1}{2}(\cos^2\theta-3\sin^2\theta\cos^2\varphi_1)+(\cos^2\theta+3\sin^2\theta\cos^2\varphi_1)+\mathcal{O}\left(\frac{k_3}{k_1}\right)\bigg] ~,
\label{eqt.IBCangle}
\end{align}
So
\begin{align}
F^{BC}(\cos\theta)=\frac{4c^3_0H^2}{m^2_Z}\frac{k_3^3}{(2\pi)^2}i\left[\frac{9}{4}-\frac{3}{4}\cos^2\theta
\right] ~,
\end{align}
\begin{align}\nonumber
	f^{BC}_1(\mu,c)
\nonumber
\xrightarrow[\mu >c>0]{\mu,|\mu-c|\rightarrow\infty}&
e^{-2 \pi (\mu-c)} (1-e^{-2 \pi  c}) 2^{-10+2 i \mu }(-\pi) {\mu ^{-1}}(2 \mu +i)^3  (2 \mu +3 i) (\mu +3 i)\times \\ \nonumber
\times & \left\{(\mu +i) (2 \mu +i) \left[(1+e^{-2 \pi  c}) \mu ^{4 i \mu } (\mu -c)^{-2 i (\mu -c)} (c+\mu )^{-2 i (c+\mu )}\right.\right.\\ \nonumber
&\hspace*{70pt}\left.-2 e^{-\pi  c} \mu ^{2 i \mu } (\mu -c)^{-i (\mu -c)} (c+\mu )^{-i (c+\mu )}\right]+\\ \nonumber
&+e^{-\pi  (\mu -c)}\cdot 2 e^{i\pi/4}(\pi \mu)^{-1/2} \left[e^{-\pi  c} (1+e^{-2 \pi  c}) (2 \mu +i)\right.\\ \nonumber
&\hspace*{70pt}\left.-\left((1+e^{-4 \pi  c})\cdot i+e^{-2 \pi  c}\cdot 4 \mu\right)\mu ^{2 i \mu }(\mu -c)^{-i (\mu -c)} (c+\mu )^{-i (c+\mu )}\right]\\
&\left.+\mathcal{O}\left(e^{-2\pi(\mu-c)}\right)\right\} ~,\\
\nonumber
\xrightarrow[c>\mu>0]{\mu,|\mu-c|\rightarrow\infty}&
e^{4 \pi ( c-\mu)}{ 2^{-10+2 i \mu }(-\pi ) \mu ^{-1}} (2 \mu +i)^3  (2 \mu +3 i) (\mu +3 i)\times\\ \nonumber
\times & \left[(1+e^{-2 \pi  \mu })^{-1}\cdot 2e^{-i\pi/4}(\pi  \mu)^{-1/2}  \mu ^{2 i \mu } (c+\mu )^{-i (c+\mu )} (c-\mu )^{i (c-\mu )}\right.\\ \nonumber
&+(1-e^{-4 \pi  \mu }) (2 \mu +i) (\mu +i)  \mu ^{4 i \mu } (c-\mu )^{2 i (c-\mu )} (c+\mu )^{-2 i (c+\mu )}\\
&\left.+(1-e^{-4 \pi  \mu })^{-1}(-i) (\mu -i) (2 \mu -i)+\mathcal{O}\left(e^{-2\pi(c-\mu)}\right)\right] ~,
\end{align}
and
$
	f^{BC}_3(\mu,c)=-f^{BC}_1(\mu,c)^* ~.
$
\paragraph{Contribution by $I^{\ ba}_{CA,(0)}$}
The contribution to \eqref{eqt.diagram}
\begin{align}\nonumber
	&\frac{4c_0^3H^2}{m_Z^2}\int \frac{\mathbf{d}^3\mathbf{p}}{(2\pi)^3}k_1^{-2}k_3^{-2}i\{[-(\mathbf{p}\cdot i\mathbf{k_1}) (\mathbf{q}\cdot i\mathbf{k_2})  - (\mathbf{p\cdot q})(i \mathbf{k_1}\cdot i\mathbf{k_2})+(\mathbf{p}\leftrightarrow\mathbf{q})]
+(\mathbf{k_1}\leftrightarrow\mathbf{k_2})\}
\\
\simeq & \frac{4c_0^3H^2}{m_Z^2} \frac{k_3^3}{(2\pi)^3}\int^{2\pi}_0 d\varphi_1\ i\left[2+(\cos^2\theta-3\sin^2\theta\cos^2\varphi_1)
\right] ~,
\label{eqt.ICAangle}
\end{align}
So
\begin{align}
F^{CA}(\cos\theta)=\frac{4c^3_0H^2}{m^2_Z}\frac{k_3^3}{(2\pi)^2}i\left[\frac{1}{2}+\frac{5}{2}\cos^2\theta
\right] ~,
\end{align}
\begin{align}
	f^{CA}_1(\mu,c)
\nonumber
\xrightarrow[\mu>c>0]{\mu,|\mu-c|\rightarrow\infty}&
e^{-\pi  (\mu - c)}\left(1-e^{-2 \pi  c}\right) 4^{-5+i \mu } \pi \mu ^{-1} (2 \mu +i)^4 (2\mu+3 i)(\mu+3 i)\times \\ \nonumber
\times&\left[-e^{i\pi/4} (\pi \mu)^{-1/2} \mu ^{2 i \mu } (\mu -c)^{i (c-\mu )} (c+\mu )^{-i (c+\mu )}+ e^{-\pi (\mu - c) } (1+e^{-2 \pi  c})\times \right.\\
& \times\left.(\mu +i) \mu ^{4 i \mu } (\mu -c)^{2 i (c-\mu )} (c+\mu )^{-2 i (c+\mu )}+\mathcal{O}\left(e^{-2\pi(\mu-c)}\right)\right] ~,\\
\nonumber
\xrightarrow[c>\mu>0]{\mu,|\mu-c|\rightarrow\infty}&
 e^{4 \pi ( c- \mu) } 2^{-10+2i \mu } \pi {\mu }^{-1} (2 \mu +i)^3 (2 \mu +3 i) (\mu +3 i)\times\\ \nonumber
\times&\left[(1+e^{-2 \pi  \mu })^{-1} 2 e^{-i\pi/4} (\pi  \mu)^{-1/2} \mu ^{2 i \mu } (c-\mu )^{i (c-\mu )} (c+\mu )^{-i (c+\mu )}\right.\\ \nonumber
&\left.+ (1-e^{-4 \pi  \mu }) (\mu +i) (2 \mu +i) \mu ^{4 i \mu } (c-\mu )^{2 i (c-\mu )} (c+\mu )^{-2 i (c+\mu )}\right.\\
&\left.+(1-e^{-4 \pi  \mu })^{-1}(-i)(\mu -i) (2 \mu -i)+\mathcal{O}\left(e^{-2\pi(c-\mu)}\right)\right] ~,
\end{align}
and
$
	f^{CA}_3(\mu,c)=-f^{CA}_1(\mu,c)^* ~.
$
\paragraph{Contribution by $I^{\ ba}_{AC,(0)}$}
The contribution to \eqref{eqt.diagram}
\begin{align}\nonumber
	&\frac{4c_0^3H^2}{m_Z^2}\int \frac{\mathbf{d}^3\mathbf{p}}{(2\pi)^3}k_1^{-2}k_3^{-2}i\{[ p^2 (i \mathbf{k_1}\cdot i\mathbf{k_2}) + (\mathbf{p}\cdot i\mathbf{k_1})(\mathbf{p}\cdot i\mathbf{k_2}) +(\mathbf{p}\leftrightarrow\mathbf{q})]
+(\mathbf{k_1}\leftrightarrow\mathbf{k_2})\} 
\\
\simeq & \frac{4c_0^3H^2}{m_Z^2} \frac{k_3^3}{(2\pi)^3}\int^{2\pi}_0 d\varphi_1\ i \left[(3 \sin ^2\theta \cos ^2{\varphi_1}+\cos ^2\theta)+4 \right] ~,
\label{eqt.IACangle}
\end{align}
So
\begin{align}
F^{AC}(\cos\theta)=\frac{4c^3_0H^2}{m^2_Z}\frac{k_3^3}{(2\pi)^2}i\left[\frac{11}{2}-\frac{1}{2}\cos^2\theta
\right]
\end{align}
\begin{align}
	f^{AC}_1(\mu,c)
\nonumber
\xrightarrow[\mu>c>0]{\mu,|\mu-c|\rightarrow \infty}&e^{-\pi  (\mu -c)}(1-e^{-2 \pi  c})  2^{-10+2 i \mu }\pi \mu ^{-1}(2 \mu +i)^3 (2 \mu +3 i)(\mu +3 i)\times  \\ \nonumber
\times&\left[e^{i\pi/4}(\pi \mu)^{-1/2} (2 \mu -i) \mu ^{2 i \mu }  (\mu -c)^{i (c-\mu )} (c+\mu )^{-i (c+\mu )} + e^{-\pi (\mu -c)} (1+e^{-2 \pi  c})\times\right. \\
&\times\left.(\mu +i) (2 \mu +i) \mu ^{4 i \mu } (\mu -c)^{2 i (c-\mu )} (c+\mu )^{-2 i (c+\mu )}+\mathcal{O}\left(e^{-2\pi(\mu-c)}\right)\right] ~,\\ \nonumber
\xrightarrow[c>\mu>0]{\mu,|\mu-c|\rightarrow \infty}& e^{4 \pi  (c-\mu)} 2^{-10+2 i \mu }\pi  \mu^{-1}(2 \mu +i)^3 (2 \mu +3 i) (\mu +3 i)\times\\ \nonumber
\times&\big[(1+e^{-2 \pi  \mu })^{-1}2  e^{-i\pi/4} \pi^{-1/2} \mu^{-1/2} \mu ^{2 i \mu } (c-\mu )^{i (c-\mu )} (c+\mu )^{-i (c+\mu )}\\ \nonumber
&+(1-e^{-4 \pi  \mu }) (\mu +i) (2 \mu +i) \mu ^{4 i \mu } (c-\mu )^{2 i (c-\mu)} (c+\mu )^{-2 i (c+\mu )}\\
&+\left.(1-e^{-4 \pi  \mu })^{-1}(-i) (\mu -i) (2 \mu -i)+\mathcal{O}\left(e^{-2 \pi(\mu-c)}\right)\right] ~,
\end{align}
and
$
	f^{AC}_3(\mu,c)=-f^{AC}_1(\mu,c)^* ~.
$
\\\\
From $F^{BC}$, $F^{CA}$ and $F^{AC}$, we have
\begin{align}
	f_\theta(\mu,c)=i\left[-\frac{3}{4}f^{BC}_1(\mu,c)+\frac{5}{2}f^{CA}_1(\mu,c)-\frac{1}{2}f^{AC}_1(\mu,c)\right]& ~,\\
	f_c(\mu,c)=i\left[\frac{9}{4}f^{BC}_1(\mu,c)+\frac{1}{2}f^{CA}_1(\mu,c)+\frac{11}{2}f^{AC}_1(\mu,c)\right]& ~.
\end{align}
\newpage
\bibliography{axioncollider_arxiv}{}
\bibliographystyle{utphys}
\end{document}